\documentclass{sig-alternate}

\usepackage[ruled,vlined,linesnumbered]{algorithm2e}
\usepackage[utf8]{inputenc}
\usepackage{subfigure}
\usepackage{mathtools}
\usepackage{multicol}
\usepackage{graphicx}
\usepackage{rotating}
\usepackage{mathptmx}
\usepackage{amsfonts}
\usepackage{balance}
\usepackage{amsmath}
\usepackage{amssymb}
\usepackage{wrapfig}
\usepackage{times}
\usepackage{color}
\usepackage{tikz}
\usepackage{url}

\usetikzlibrary{decorations.pathreplacing,shapes}

\newcommand{\eat}[1]{}

\newenvironment{packed_enum}{
\begin{enumerate}
  \setlength{\itemsep}{0pt}
  \setlength{\parskip}{0pt}
  \setlength{\parsep}{0pt}
}{\end{enumerate}}

\SetKwRepeat{Do}{do}{while}

\title{Vers\~o foli\~o: Diversified Ranking for Large Graphs with Context-Aware Considerations}

\numberofauthors{1} 
\author{
\alignauthor
George Tsatsanifos\\
\affaddr{HomeGrown Research Labs}\\
\affaddr{Omirou 39, Agios Dimitrios,}\\
\affaddr{172 36, Athens, Greece}
}

\begin{document}

\maketitle

\begin{abstract}
This work is pertaining to the diversified ranking of web-resources and interconnected 
documents that rely on a network-like structure, e.g. web-pages. A practical example of 
this would be a query for the $k$ most relevant web-pages that are also in the same time 
as dissimilar with each other as possible. Relevance and dissimilarity are quantified using an 
aggregation of network distance and context similarity. For example, for a specific 
configuration of the problem, we might be interested in web-pages that are similar with 
the query in terms of their textual description but distant from each other in terms of 
the web-graph, e.g. many clicks away. 
In retrospect, a dearth of work can be found in the literature addressing this problem 
taking the network structure formed by the document links into consideration. In addition, 
the vast majority of the approaches that have been proposed in the literature leverage 
greedy heuristics relying on certain measures that capture only some aspect of the problem. 
Arguably, this is not enough. 

In this work, we propose a hill-climbing approach that is seeded with a document 
collection which is generated using greedy heuristics to diversify initially. More 
importantly, we tackle the problem in the context of web-pages where there is an 
underlying network structure connecting the available documents and resources. This 
is a significant difference to the majority of works that tackle the problem in 
terms of either content definitions, or the graph structure of the data, but never 
addressing both aspects simultaneously. To the best of our knowledge, this is the 
very first effort that can be found to combine both aspects of this important problem 
in an elegant fashion by also allowing a great degree of flexibility on how to 
configure the trade-offs of (i) document relevance over result-items' dissimilarity, 
and (ii) network distance over content relevance or dissimilarity. Last but not 
least, we present an extensive evaluation of our methods that demonstrate the 
effectiveness and efficiency thereof.
\end{abstract}

\vspace{-8pt}
\category{H.3.3 }{Information Search and Retrieval}{Search Process}
\vspace{-8pt}

\section{Introduction}
\label{sec:intro}

Result diversification has lavished great scientific interest in recent years for it 
addresses a variety of problems simultaneously. First, modern information retrieval 
research focuses on result diversification as a means of counteracting intrinsic IR 
problems, like the over-specialization problem of retrieving too homogeneous results.
For instance, for information queries, users reading through a list of relevant but 
redundant pages quickly stop as they do not expect to learn more. Therefore, new search 
techniques had to be improvised in order to ameliorate user satisfaction and decrease 
query abandonment. Second, diversification reduces the risk that none of the returned 
results satisfies a user's query intent. For example query term ``\emph{Apache}'' could 
pertain to the well-known software foundation, or the Native American tribe just as well. 
Similarly, among the results for ``\emph{Python}'' we could find information about a 
reptile specie, a programming language, or even the Monty Python, a comedy group. 
Therefore, by focusing on just one genre of information, we could easily fail to 
answer the query properly, or even danger to be completely irrelevant. Also, among 
the most frequent examples found in the literature for ambiguous and exploratory 
queries ``\emph{flash}'' and ``\emph{jaguar}'' are the most typical for they can 
represent completely different things, e.g. cars, felines, etc. Ambiguity is a 
common and usual problem, a solution to which can be result diversification. Third, 
enabling diversity tacitly facilitates (near) duplicates elimination.

Besides, conventional information retrieval models presume that given a similarity measure 
with different degrees of relevancy, the relevance of a document is independent of the 
relevance of other documents. In reality however, this \emph{independence relevance} assumption 
rarely holds; the utility of retrieving one document may depend on which documents the user 
has already seen. Regarding this complementary aspect of diversification, it is in general 
insufficient to simply return a set of relevant results, since correlations among them are 
also important. More specifically, documents should be selected progressively according to 
their relevance and also in terms with the documents that come before it. For example, a 
system that delivers documents that are relevant and novel must be able to identify relevant 
documents that are dissimilar to the previously delivered documents, in the sense of comprising 
new information. The main challenge in diversification lies in that these two important goals 
are contradictory to each other. 

\begin{figure*}[htb]
 \centering
 \includegraphics[width=.48\textwidth]{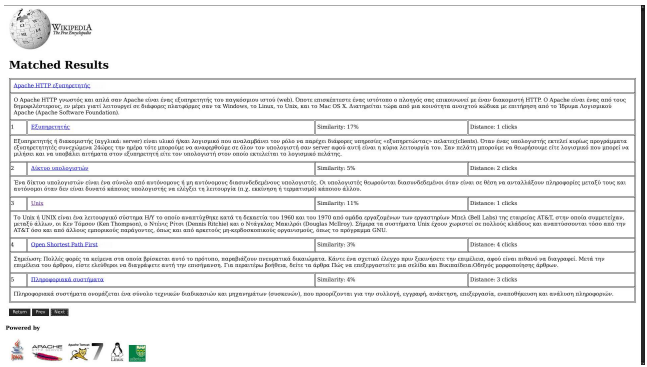}
 \hspace{15pt}
 \includegraphics[width=.48\textwidth]{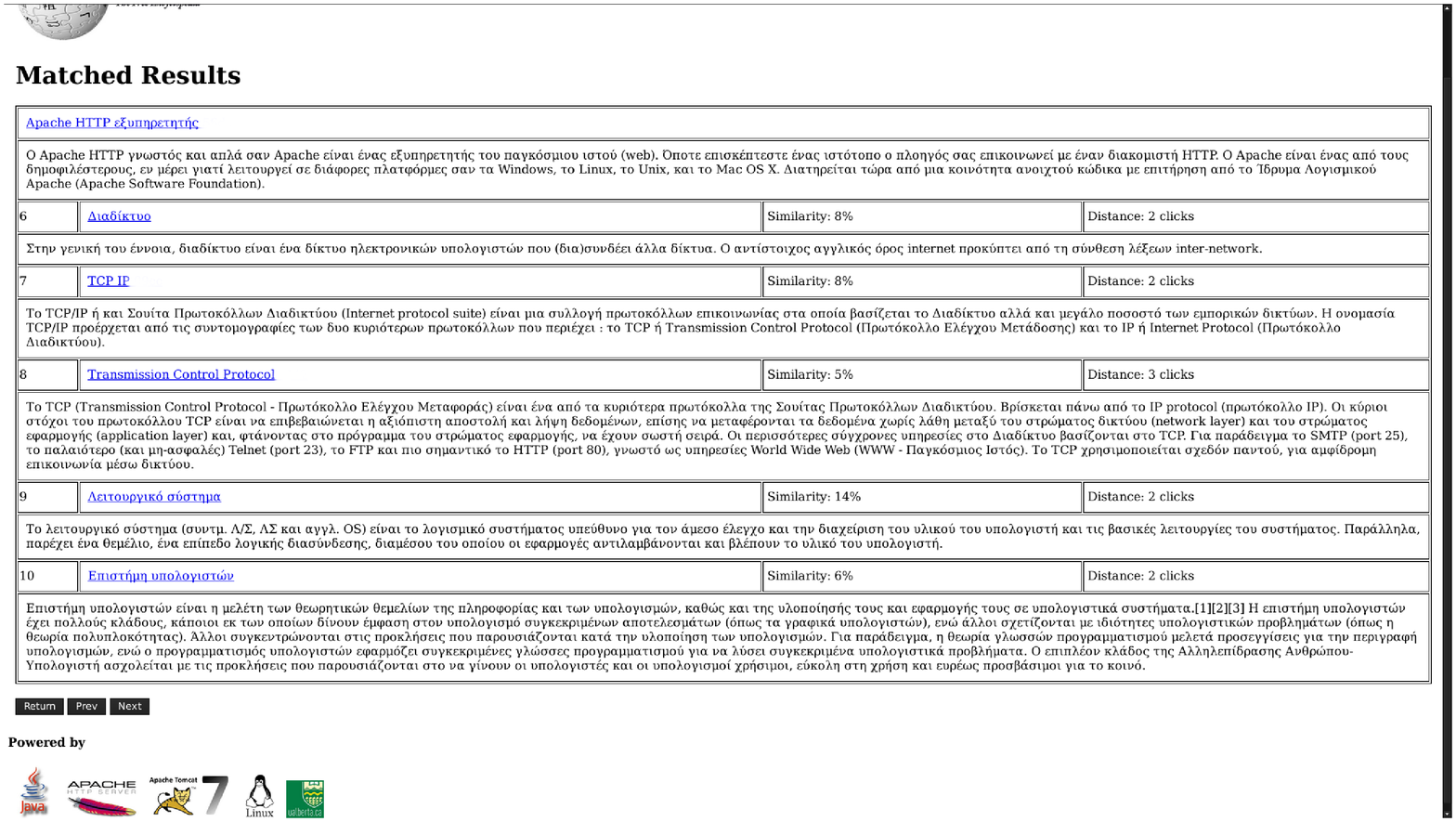}
 \caption{Diversified result for query ``\emph{Apache}'' over the Greek collection of Wikipedia articles. 
          The ``\emph{Apache HTTP server}'' article was chosen as the center for this search because of 
          its title. Relevant articles representing different fields are returned in the result.}
 \label{fig:apache}
\end{figure*}

In essence, the goal of diversification is to identify a list of items that are dissimilar with 
each other, but nonetheless relevant to the user's search criteria. In other words, the diversity 
is defined by how much each item of the result-set differs from the others in terms of their 
attribute values. In effect, diversification is a multifaceted problem with many practical and 
theoretical extensions. A profound and rich link with combinatorics and approximation is exploited 
in research works where diversification is performed through heuristics and greedy algorithms mostly.
This work focuses on diversified ranking of web-resources and documents that rely on a 
\emph{network-like structure}, e.g. web-pages. A typical example of this problem would 
be a query for the $k$ most relevant web-pages that are also in the same time as dissimilar 
with each other as possible. This is an important problem with deep rooted theoretical 
extensions and even more practical applications. It is a NP-hard problem and the fact 
that we address it in the context of data representing interconnected information, adds 
not only to the complexity of the problem, but also to the value of an effective and 
efficient solution that is capable of returning a near-optimal solution within a few seconds.

More specifically, there are two approaches proposed in this paper used in 
combination, one after the other, in an effort to maximize their effect. The 
first relies on greedy heuristics, whereas the second leverages a hill-climbing 
approach. According to the former, we compute $k_g$ diversified sets starting 
from the object in the network that is most similar to the query, given a 
metric that smooths network distance with context similarity, and we incrementally 
append to the set the object which when added to the result constructed so far, the 
rank of the outcome is better than it would be if we had added any other object.
This process is repeated as many times as the desired number of elements. On the 
other hand, the latter approach, takes as input the $k_g$ greedily built 
diversified sets, and iteratively refines them by removing certain elements and 
replacing them with more preferable choices in order to produce better diversified 
sets. Each time we keep only the best $k_c$ derived sets we have retrieved so far, 
and we work on that collection of sets as a means to explore the overall search space. 

What is more, we prove our concept through a prototype that we launched on-line.
It operates over multi-lingual collections of the Wikipedia dataset but other 
collections of web-pages can also be used as soon as the appropriate models are 
extracted, e.g. web-graph, a vector space model (based on tf-idfs in our case), 
etc. In Figure \ref{fig:apache}, we depict the results for ``\emph{Apache}''.
Among those one can find links to articles up to 4 clicks away while still being 
relevant to the ``\emph{Apache HTTP server}'', the article which was chosen to 
serve as the query center for our search because of its title for it matches the 
query better than other articles. Among the top results ``\emph{Server}'' and 
``\emph{Unix}'' are the closest articles that are also very similar to the query 
center in the same time. Of course, other close neighbors could have been selected, 
but we put special emphasis on the fact that those are quite diverse with each other 
as they belong to different categories, tacitly addressed though a model that relies 
on content-based definitions. When we examine the remaining results (not directly 
connected to the query center) we see links to other articles like ``\emph{Computer 
Networks}'', ``\emph{Information Systems}'', ``\emph{Internet}'', ``\emph{Operating 
Systems}'', and ``\emph{Computer Science}'', each contributing to the result by 
representing a different field or thematic area, tackling this way directly the 
over-specialization problem, the main challenge addressed in this paper. Other 
notions, like the page-rank of each web-page could easily be incorporated in our 
rank functions as a way to promote more important pages, exactly like we do with 
the relevant ones. All different criteria are smoothed according to the user 
preferances via an elegant interface.

In summary, the contributions of this paper are the following:
\begin{packed_enum}
  \item	We present \texttt{verso}, an algorithm which given a query object $q$, 
        a set of interconnected objects $S$, and a ranking function $f$, it retrieves 
        the \emph{optimal} single object, which when added in $S$ it is better ranked 
        in terms of both network characteristics and textual descriptions of its 
        elements, than it would be if any other point had been inserted.
  \item	We propose a paradigm that combines a method based upon greedy heuristics with 
        an approach leveraging a hill-climbing methodology to exploit \texttt{verso}, 
        so as to construct diversified sets of arbitrary cardinality.
  \item	We show through an extensive experimental evaluation that our paradigm clearly 
        outperforms the \texttt{BestCoverage} method proposed in \cite{KucuktuncSKC13} 
        in terms of response times, result quality, and memory requirements, as well.
  \item We developed in Java and made available on-line a prototype implementing our method.
\end{packed_enum}

As a final note, the name of our paradigm ``\emph{vers\~o foli\~o}'' translates from Latin to 
English as ``\emph{the turned side of the page}''.
The remainder of this paper is organized as follows: Section \ref{sec:related} discusses contemporary 
diversification techniques over different types of data. A formal definition of the problem we study 
here is presented in Section \ref{sec:problem}. Our paradigm is presented in Section \ref{sec:retrieval}.
Section \ref{sec:exp} presents an extensive evaluation of our scheme. Finally, we conclude and summarize 
our contributions in Section \ref{sec:concl}.

\vspace{-5pt}
\section{Related Work}
\label{sec:related}

In this section, we present an overview of the literature on result diversification over two different fields.

\subsection{Result Diversification for Structured Data}

The first part of this section describes the various definitions of the result diversification problem 
proposed in the research literature and classifies the algorithms proposed for locating diverse items. 
A complete survey on the state of the art can be found in \cite{DrosouP10}.

In the absence of any explicit knowledge of user intent, three main approaches are employed for result 
diversification: (i) \emph{content} based, for differentiated result items in terms of their attribute 
values, (ii) \emph{novelty} based, when promoting items that contain new information compared to those 
ranked higher, and (iii) \emph{coverage}, for including items so as to cover as many categories as possible.

First and foremost, content-based definitions aim at maximizing the aggregated distance between 
any pair of objects in the result-set. Arguably, the most influential work on diversification 
is that of Maximal Marginal Relevance (MMR) \cite{CarbonellG98}. In their seminal work, the 
authors introduced a trade-off between diversity and the relevance of search results through 
the combination of two similarity functions shown in Equation \ref{eq:mmr}. One for measuring 
the similarity among a document collection $S$, and the other the similarity between query $q$ 
and each document. In addition, parameter $\lambda$ controls the degree of trade-off.
\begin{equation}
  \max_{x \in S} \lambda d(q,x) - (1-\lambda) \min_{x,y \in S} d(x,y) \label{eq:mmr}
\end{equation}

Moreover, \cite{GollapudiS09} is another conspicuous work that follows an axiomatic approach to 
develop a set of natural axioms that a diversification system is expected to follow. A variety of methods 
are discussed, such as semantic distance-based schemes, taxonomy-based categorical distance schemes, and 
the advantages and disadvantages thereof. On the other hand, \cite{VeeSA09} focuses on diversification 
over structured data to propose an appropriate ranking-similarity function for a given ordering of the data attributes. 
An interesting approach is presented in \cite{Haritsa09} that extends the $k$ nearest neighbors problem 
with providing a minimum distance threshold among the points of the result-set in order to guarantee diversity. 
In a more recent work \cite{topKsigmod}, the authors adopt an approach based on Voronoi cells in order to retrieve 
the object maximizing a novelty function. More specifically, items that already constitute the answer are appointed 
as cell centers. Then, all objects that happen to be placed on the vertices and the edges of the formed Voronoi 
diagram are examined and the best of them is inserted into the answer-set. Another similar approach is 
proposed in \cite{marios} which however requires knowing in advance all the objects to compute the objective 
function.

Novelty can be seen as a means to avoid redundancy and (near) duplicates, and also as a means to impart 
additional information to the result-set; whereas diversity is viewed as the need to resolve ambiguity. 
Adhering to this principle, queries and objects are treated as sets of ``information nuggets'' in 
\cite{ClarkeKCVABM08}, while relevance is a function of the nuggets contained in possible interpretations 
of the query and the result-set. In another line of work, \cite{ZhangCM02} empowers adaptive filtering 
systems to distinguish novel and redundant items. Such systems identify items that are similar to 
previously delivered ones in having the same topic, but also dissimilar to them, in the sense of 
containing novel information.

Conversely, \cite{XuY08} suggests from the user point of view that ``novelty seeking'' is not equivalent 
to  ``diversity seeking'', and that the novelty preferences of individual users are directed towards finding 
more information on specific subtopics of interest, rather than an undirected quest for any new information 
(as in the coverage based definitions). 
In \cite{AgrawalGHI09} this approach is augmented with taking into account the relative importance of different 
nuggets (as distribution categories), and the fact that different documents containing the same nugget may satisfy 
the user to different extent. 
On the other hand, \cite{ZhaiCL03} suggests models used in traditional IR but seeks to modify the ranking so as 
to include documents relevant to several subtopics in terms of a taxonomy.

Furthermore, the diversification problem has been shown to be NP-hard, since it is directly related to the 
$p$-dispersion problem. To elaborate, the objective of the $p$-dispersion problem is to choose $p$ out of $n$ given 
points, so that the minimum distance between any pair of chose points is maximized. Quite often however, the objective 
function is the average distance of any two points. The authors in \cite{GollapudiS09} consider different formulations 
of the problem, such as \emph{min-max diversification}, \emph{min-sum}, and a \emph{mono-objective} formulation of the 
problem. Additionally, the combination of these criteria has been studied in \cite{ZhangH08} as an optimization problem. 
Besides, due to the increased complexity of the task, most efforts rely on approximation. Many heuristic algorithms have 
been proposed in the diversity discourse and have been employed for solving variations of the problem. These heuristics 
can be classified into two main categories: (i) greedy, and (ii) interchange (or hill-climbing).

The majority of the proposed algorithms adopt a \emph{greedy} approach for being intuitive and relatively fast. A 
straightforward methodology is proposed in \cite{AgrawalGHI09}, according to which, given the set of the top-$k$ most 
relevant documents to a query, we re-order them in such a way that the objective function of the model 
is optimized. Other simplistic solutions rely on post-processing returned items to identify those that differ in their 
attribute values. In the context of recommender systems, \cite{ZieglerMKL05} presents a conspicuous greedy paradigm, 
according to which the result-set is initialized with the most relevant recommendation. Then, recommendations that are 
not part of the answer are sorted in terms of their relevance to the query and their minimum distance from any item of 
the answer set. Now, the rank of each recommendation is a linear combination of its position in the two lists. At each 
iteration, we append to the result-set the recommendation with the minimum rank, until the answer consists of $k$ 
recommendations. Another similar approach is proposed in \cite{YuLA09}. Starting with the most relevant item, it adds 
the next most relevant item if and only if that item is far away (compared to a distance bound) from the existing items 
in the set, until we have included $k$ items. A similar algorithm is proposed in \cite{Haritsa09} in the context of 
spatial databases.

Interchange heuristics start with a random result-set and then iteratively attempt to improve 
it by swapping an item in the solution with another item that is not part of the answer. At each iteration, possible 
swap candidates are the ones that ameliorate diversity the most. \cite{LiuSC09} constitutes such an effort in the 
context of (semi-) structured data. Another additional interchange heuristic is presented in \cite{YuLA09}, where 
starting with the $k$ most relevant items; at each iteration, the item that contributes the least to the diversity, 
is substituted by the most promising item that is not part of the answer set. This sequential procedure terminates 
when there are no more items with better diversity than a given threshold.
%
In \cite{ripple} an approach that leverages both methodologies is proposed in the context of structured 
peer-to-peer networks. In \cite{phdworkshop}, we propose similar diversification principles for spatial 
data indices. In \cite{disc1,disc2}, DisC is proposed more recently for computing diversified results of 
variable approximation by processing a hierarchical tree structure accordingly using a top-down method.

\subsection{Diversified Search over Graph Data}

Most of the works that can be found in the literature on the retrieval of diversified 
sets of vertices use non-negative monotone submodular set functions for ranking. The 
main advantage that comes with this is that the near-optimality is easily achieved within 
$(1-\frac{1}{e})$-approximation from the optimal solution using greedy heuristics. A 
theoretical analysis about the approximation guarantees can be found in \cite{NemhauserWF78, 
CalinescuCPV11}. However, our work considers bicriteria ranking functions that are not 
necessarily submodular, but instead emphasize on a more general family of the Maximal 
Marginal Relevance (MMR) ranking functions, and therewith, near-optimality is much harder 
to be achieved. Besides, the problem we tackle in this paper is NP-hard and the network 
structure of the data we delve with sets even more challenges, practical and theoretical, 
as well.

In \cite{LiY13} the authors propose a diversified ranking measure on large graphs 
that captures the relevance to the query and the dissimilarity of the elements 
in the result. They formulate the problem as a submodular set function optimization 
problem. Their approach relies on a notion of expansion, according to which a set 
of nodes with a large expansion ratio implies that the nodes are dissimilar to 
each other, in a sense that they do not share the common neighbors in a graph.
Subsequently, the larger expansion ratio the set of nodes has, the better 
diversity we achieve. Their ranking function is designed accordingly in a way 
that compromises the personalized PageRank scores over the ranking results, 
compromising this way their relevance with their expansion ratio. Nevertheless, 
the ranking function does not consider the ordering of the list items, in a 
sense that a permutation of the elements of the result would yield exactly the 
same score. On the other hand, since our scheme relies also on greedy heuristics, 
each new result item is selected for it produces an augmented set that achieves a 
better score than any other set derived by adding an alternative node to the 
set. Hence, previous selections play a dominant role on which node will be 
included next.

The authors in \cite{KucuktuncSKC13} propose a measure that also corresponds 
to a submodular set function, they call \emph{expanded relevance} and is used 
to compute the coverage of the ``interesting'' part of the graph. However, they 
argue that treating the diversification problem as a bicriteria optimization 
problem with a relevance measure that ignores diversity, and a diversity measure 
that ignores relevancy can turn out problematic. Towards this end, they propose 
a greedy algorithm which at each iteration considers including in the result-set 
a vertex among $k \delta^\ell$ ranked vertices, where $\delta$ the average fan 
degree and $\ell$ a system parameter dictating nodes from how many hops away we 
should consider including in the result. The best among those nodes is selected 
to augment the result and the impact of its distance $\ell$ neighbors can be 
adjusted. On the downside, we reckon that this is somehow limiting as each time 
only vertices that are up to $\ell$ hops away from the tentative result are 
considered, and no object in distance greater than $k \ell$ from the query 
center is ever examined, even though it could constitute a much better option.

The method proposed in \cite{HeTMS12} assumes that all pairwise distances are 
pre-computed and made available in a square matrix and given a query the relevance 
of each vertex is computed. Given these parameters, the proposed algorithm greedily 
appends to the result-set the element that maximizes a relevance factor that also 
captures clustering properties, e.g., a relevant object that is close to other 
relevant items forming a cluster this way is ranked higher, and penalizes the 
items that are close to the already selected ones. 

GrassHopper \cite{ZhuGGA07} incrementally selects each time the best ranked vertex 
and turns it into a sink node. Leveraging random walks, the algorithm converges when 
all sinks are traced, and estimates the ranks with the number of visits to each node 
before convergence. It uses a matrix inversion to find the expected number of visits; 
nevertheless, inverting a sparse matrix makes it dense, which is not practical for the 
large and sparse graphs that the research community is interested in, e.g. web-graphs, 
social networks, etc.

In another line of work, Dragon \cite{TongHWKL11} relies on greedy heuristics and they 
propose a submodular set function for ranking that combines relevance with the ``Google 
matrix'' \cite{ilprints422} in an effort to capture the strength of the connection between 
any two elements. Similarly, DivRank \cite{MeiGR10} adjusts a transition matrix based on 
the number of visits to the vertices using a random walk variant called vertex-reinforced 
random walks that assumes that there is a link for all vertices returning back to the node 
itself, which is followed with some probability. The method ensures that the highly ranked 
nodes collect more value over the iterations, resulting in a rich-gets-richer phenomenon.

\vspace{-5pt}

\begin{figure*}
\begin{eqnarray}
  \sigma_q (S) = \frac{\lambda}{N} \sum_{u \in S} \{ \alpha d^\text{graph}(q,u) + (1-\alpha) d^\text{text}(q,u) \}
  + \frac{1-\lambda}{N(N-1)} \sum_{v \in S} \sum_{\substack{w \in S\\ v \prec w}} \{ \beta d^\text{graph}(v,w) + (1-\beta) d^\text{text}(v,w) \} \label{eq:avg} \\
  \sigma_q (S) = \lambda \max_{u \in S} \{ 
                 \underbrace{\alpha d^\text{graph}(q,u) + (1-\alpha) d^\text{text}(q,u)}_{d^{\text{rel}}(q,u)}
                 \}
                 + (1-\lambda) \min_{\substack{v,w \in S\\ v \prec w}} \{ 
                 \underbrace{\beta d^\text{graph}(v,w) + (1-\beta) d^\text{text}(v,w)}_{d^{\text{diss}}(v,w)}
                 \} \label{eq:max}
\end{eqnarray}
\end{figure*}

\section{Problem Specification}
\label{sec:problem}

The problem we study in this paper can be described as following: Given a 
\emph{directed} graph $G(V,E)$, where $V$ and $E$ its sets of vertices and 
edges respectively, we want to find the best set $S$ of $k$ vertices that 
minimizes either of the following Maximal Marginal Relevance (MMR) ranking 
functions from Equations \ref{eq:avg} or \ref{eq:max} for a given query 
object, say $q$. Henceforth, we assume that each vertex $u$ in the network 
is associated with a textual description $t_u$, to which we may also refer 
as the context of vertex $u$. By definition, the combination of the two, a 
vertex from $V$ along with its textual description, constitutes an object.

The association becomes now clear. According to the adopted model each vertex 
represents a web-page, its \emph{directed} edges correspond to its links, and 
its textual description to its content.
Our model is versatile and flexible enough to accomodate any measure of similarity 
between two textual descriptions. For our purposes, we employ the cosine similarity, 
and thus, we define the distance of two vertices $u$ and $v$ given their contexts 
$t_u$ and $t_v$, respectively, as $d^\text{text}(u,v) = 1- \frac{\langle t_u \cdot 
t_v \rangle}{|t_u| |t_v|}$, and is smoothed accordingly by the network distance in 
an affine combination using special variables dedicated to that purpose. To elaborate, 
parameter $\lambda$ takes values in $[0,1]$ in Equations \ref{eq:avg} and \ref{eq:max}, 
and denotes the desired trade-off between relevance and result dissimilarity. Likewise, 
$\alpha$ and $\beta$ also take values within the same interval and stand for the 
compromise between the network distance and context similarity of any pair of objects. 
Moreover, each of the two ranking functions adopts different criteria on how to quantify 
collectively the different characteristics and properties of an input-set and compromise 
the relevance of its elements with their degree of dissimilarity. More importantly, the 
formulation of the problem we study here is novel, in a sense that this is the first work 
to compromise the structure of the network with the textual descriptions of its elements, 
allowing in the same time a great degree of freedom to define exactly how much of those 
features should contribute to the rank of the result, by either configuring the $\alpha$ 
parameter for its relevance, or $\beta$ for the desired pairwise dissimilarity of the 
result items. For example, for a specific configuration of the problem, we might be 
interested in web-pages that are similar with the query in terms of their textual 
description only, and hence we set $\alpha=0$, but as much distant from each other as 
possible in terms of their network properties, and we therefore set $\beta=1$. In 
addition, we can specify, how much we want relevance and result dissimilarity to 
contribute in the ranking process using the $\lambda$ parameter. For example, we might 
want to promote sets that congregate better representatives of the domain than relevant 
ones for low $\lambda$-values, and vice versa for high. In Equations \ref{eq:avg0} and 
\ref{eq:max0} we show the form that Eqs.~\ref{eq:avg} and \ref{eq:max} take when we are 
interested in relevance according to web-pages' content and also want to put the items of 
the result as far as possible from each other. 

Let's have a closer look now on how the distance functions are used and it will become 
clear how objects that have been added previously affect the selection of the succeeding 
elements. We note that just like the web-graph, the conceptual graph considered in our 
paradigm is directed. The first factor of our ranking functions, and we note with 
$d^{\text{rel}}(.,.)$, aggregates the relevance of the elements in the result. Regarding 
nodes' textual descriptions the cosine similarity is a symmetric function. Nonetheless, 
in terms of network distance, we aggregate the distances from the query center $q$ to 
each of the result items, since we are interested in the distance in clicks from the 
query center to each of the result-items, and not the opposite. Occasionally, we could 
aggregate for each result item the distances to the query center as well as the distances 
from it, so as to produce a more symmetric effect. Of course, one could argue that web-pages 
that are close to each other from both directions, rather than just one, could be more 
related to each other, but this is just a convention regarding the rank function. Likewise, 
for the second factor of our ranking functions, and we note with $d^{\text{diss}(.,.)}$, 
we use the distance from the \emph{preceding} result items in order to quantify their 
diversity. This is important as the user skims through the result items and expects to 
find more diverse elements, starting from the most relevant ones. 
In principle, web-pages possibly many hops away from the query center are examined 
each time, but the best ranked are processed first in a best-first manner, according 
to their potential to contribute to the final result. Specific variable thresholds can 
guarantee how far we should expand our search from the query center $q$ and the other 
elements of the tentative result, so as to achieve the desired level of diversification, 
always considering their context similarities, as well.

\vspace{-10pt}

\begin{figure}[hbt]
\begin{eqnarray}
  \sigma_q (S) = \frac{\lambda}{N} \sum_{u \in S} \{ d^\text{text}(u,q) \}
  + \frac{1-\lambda}{N(N-1)} \sum_{v \in S} \sum_{\substack{w \in S\\ v \prec w}} \{ d^\text{graph}(v,w) \} \label{eq:avg0} \\
  \sigma_q (S) = \lambda \max_{u \in S} \{ d^\text{text}(u,q) \}
  + (1-\lambda) \min_{\substack{v,w \in S\\ v \prec w}} \{ d^\text{graph}(v,w) \} \label{eq:max0}
\end{eqnarray}
\end{figure}

\vspace{-10pt}

Furthermore, the returned solutions are different depending on which of the criteria we 
put emphasis. For example, when processing structured data, by averaging the properties 
of the items in the result, their distribution seems to fit the data distribution, with 
many of those tuples appearing within different clusters of data mostly. On the other hand, 
for the latter type of aggregation, the points would ideally form a constellation around 
the query center for a given relevance (max distance from the query), lying almost equidistant 
from each other on the fringes of a circle having a radius analogous to the least allowed 
relevance. Depending on the expected size of the result and the trade-off between relevance 
and result dissimilarity, we should retrieve the solutions that somehow resemble that 
formation. Last but not least, the very essence of the problem becomes completely different 
when the network structure is involved. To the best of our knowledge, this is the first work 
to combine both aspects of this important problem in an elegant fashion that allows a great 
degree of flexibility on how to configure the trade-offs of relevance over result-items' 
dissimilarity and network distance over context (dis)similarity.

\vspace{-5pt}
\section{Computation of Diversified Sets}
\label{sec:retrieval}

We provide in this section a top-down presentation of our paradigm. 
At the top level we have Alg.~\ref{algo:pipeline} for combining a 
method leveraging greedy heuristics with a hill-climbing based 
approach. The latter is seeded with the solution of the greedy 
algorithm. In particular, we seed the interchange algorithm with 
$k_g$ instances of $n$ elements each, so as to produce $k_c$ even 
more diversified sets, and eventually opt for the best to return, 
approaching this way as much as possible the true optimal solution. 
In effect, it is similar to performing many searches from multiple 
points towards the optimal destination.

In Algorithm \ref{algo:interchange}, we present our hill-climbing method, 
which starting from a collection of sets, it iteratively refines them one 
by one with respect to their rank. First, we create all possible subsets 
of cardinality equal to $n-1$ each in lines 4--15, and then, insert them 
in a min-heap that is dedicated to storing them in such a way that it 
would allow us to process them according to how promising they are to be 
part of the final solution. This is what happens in lines 16--37, where 
with each iteration we take the subset that is ranked highest (line 17), 
and from then on, we iterate through its possible $i$-th addendum objects 
(lines 19--37) in terms of their eligibility to complement the examined 
subset, starting from the optimal\footnote{We describe towards the end of 
this section how the single optimal such object using our \texttt{verso} 
method for complementing the rest of the set is retrieved.} such element. 
Notably, even though the outcome is not the globally optimal solution, and 
neither the examined subset is necessarily part of it, the first encountered 
object is the best element overall to insert in the subset (line 22) so as 
to achieve the best possible score while keeping that particular subset 
composition intact. In turn, the one encountered in the next iteration of 
the loop is the second best, and so on. Then, in lines 23--29, we create 
$n-1$ new possible subsets given the new composition congregating that 
particular annex element. However, not all of them will be considered in 
the future processing steps. With the checks in lines 26 and 27, only the 
ones that are better ranked than the subsets they were derived from will be 
inserted in the heap where we maintain the subsets to be processed later on. 
Next, in lines 30--37, we check whether the newly formed candidate set should 
be inserted in the result-heap. This happens if it outranks any of the candidates 
retrieved so far. But if this is not the case, we can proceed with examining 
the next subset, since the succeeding possible replacement would make an even 
worse set, and therefore, there is no point in forming those sets to discard 
them eventually. Finally, this process is terminated when there are no more 
subsets to process.

\begin{algorithm}[!ht]
  refined = interverso($q$,$R$,greeverso($q$,$n$,$R$,$k_g$),$k_c$)\;
  \While{refined.size()$>1$}{
    refined.pop()\;
  }
  \Return refined.pop()\;
  \caption{diversify ($q, n, R, k_g, k_c$): Combining both schemes, greedy and interchange, 
                            using the solutions of the former to feed the latter.}
  \label{algo:pipeline}
\end{algorithm}

\begin{algorithm}[!hb]
  candidates = \textbf{new} MaxHeap()\;
  subsets = \textbf{new} MinHeap()\;
  iterators = \textbf{new} Map()\;
  \ForEach{seed \textbf{in} seeds}{
    \ForEach{$s$ \textbf{in} seed}{
      subset = \textbf{new} Set(seed)\;
      subset.remove($s$)\;
      subsets.push(subset)\;

      iterators.put (subset, \textbf{new} GraphDivIterable($q$,subset,$R$).iterator())\; \label{algo:interchange:iterator}
    }

    \If{candidates.size()$ < k_c$}{
      candidates.push(seed)\;
    }\Else{
      \If{$\sigma_q$(seed)<$\sigma_q$(candidates.peek())}{
        candidates.pop()\;
        candidates.push(seed)\;
      }
    }
  }

  \While{\textbf{not} subsets.isEmpty()}{
    subset = subsets.pop()\;

    replacements = iterators.get (subset)\;
    \While{replacements.hasNext()}{
      replacement = replacements.next()\;

      augmented = \textbf{new} Set(subset)\;
      augmented.insert (replacement)\;

      \ForEach{$s$ \textbf{in} subset}{
        newsubset = \textbf{new} Set(augmented)\;
        newsubset.remove(s)\;

        \If{$\sigma_q$(newsubset) $< \sigma_q$(subset)}{
          \If{candidates.size()$ < k_c$ \textbf{or} $\sigma_q$(newsubset) $< \sigma_q$(candidates.peek())}{
            subsets.push (newsubset)\;
            iterators.put(newsubset,\\
              replacements.replace(s,replacement))\; \label{algo:interchange:replace}
          }
        }
      }

      \If{candidates.size() $ < k_c$}{
        candidates.insert (augmented)\;
      }\Else{
        \If{$\sigma_q$(augmented) $ < \sigma_q$(candidates.peek())}{
          candidates.pop()\;
          candidates.insert (augmented)\;
        }\Else{
          \textbf{break}\;
        }
      }
    }
  }
  \Return cadidates\;
  \caption{interverso ($q$,$R$,seeds,$k_c$): Hill-climbing diversification algorithm.}
  \label{algo:interchange}
\end{algorithm}

In Algorithm \ref{algo:greedy}, we present our method that relies on 
greedy heuristics to generate $k_g$ diversified sets. Each of them is 
initiated with a vertex which is similar to the query center according 
to an aggregation of their network distance and context relevance dictated 
by the selected ranking function. The selection of those vertices that 
are similar to the query takes place in lines 5--17, and these initial 
sets of just one element at first will be processed in terms of their 
potential in lines 26--47, and will be incrementally augmented with well 
diversified objects that are greedily selected while taking into consideration 
the preceding elements of the set. In particular, we examine in lines 
29--47 each set that is not completely filled yet, and we create new 
ones with the same first elements but one additional element which is 
retrieved in line 30 according to its eligibility to complement the 
partial result, and is therefore placed at the next available position. 
Then, we choose which of the fully formed derived sets of $n$ elements 
to keep in lines 34--47, if they outrank any of the top-$k_g$ tentative 
results we have retrieved so far. Otherwise, we keep in lines 44--47 
only the best $k_g$ such incomplete sets with an addendum at the end 
for future processing in the following runs of the iteration. To 
elaborate on the usage of the iterator in lines 24, 28, 30, 31, 34, 
it allows us to browse the vertices in order of their suitability to 
complete the partial set in the best possible way. Their retrieval 
is based on Alg.~\ref{algo:singlepoint} and is innovative in a sense 
that it combines the context of the vertices along with their network 
distance from the items of the examined set. The iterator's \texttt{expand($s$)} 
invocation in line 34 allows us to take advantage of the graph search 
we have conducted so far, instead of creating a newly initialized 
iterator instance and perform again a significant portion of the 
search we have already done at an earlier stage. Thereby, we exploit 
the state of the previous iterator by copying its reusable elements 
into a new instance, and also add the necessary information, e.g. 
distance-heap, score-heap, etc., so as to accommodate the search 
around the new result item $s$.

\begin{algorithm}[!ht]
  similar = \textbf{new} MaxHeap()\;
  paths = \textbf{new} MinHeap()\;

  paths.push($(q,q),0$)\;
  marked.insert($q$)\;

  \While{\textbf{not} paths.isEmpty()}{
    $(q,u)$ = paths.pop()\;

    \If{$u \in R$}{
      \If{similar.size() $<k$}{
        similar.insert($u,\sigma_q$(u))\;
      }\Else{
        \If{$\sigma_q$(u)$ < \sigma_q$(similar.peek())}{
          similar.pop()\;
          similar.push($u,\sigma_q$(u))\;
        }
      }
    }
    \ForEach{adjacent node $v$ to $u$}{
      \If{\textbf{not} marked.contains($v$) \textbf{and} 
          (similar.size()$ < k_g$ \textbf{or} $(q,u)$.weight+$(u,v)$.weight $< \sigma_q$(similar.peek()))}{
        paths.push($(q,v),(q,u)$.weight+$(u,v)$.weight)\;
        marked.insert($v$)\;
      }
    }
  }

  \ForEach{$s$ \textbf{in} similar}{
    $S$ = \textbf{new} Set()\;
    $S$.insert($s$)\;
    candidates.add($S$)\;
  }

  iterators = \textbf{new} Map()\;
  \ForEach{candidate \textbf{in} candidates}{
    iterators.put(candidate, \textbf{new} GraphDivIterable($q$,candidate,$R$))\; \label{algo:greedy:iterator}
  }

  result = \textbf{new} MaxHeap()\;
  \While{\textbf{not} candidates.isEmpty() \textbf{and} candidates.first().size()$ < n$}{
    $S$ = candidates.removeFirst()\;

    iterator = iterators.get($S$)\;

    counter = 0\;
    \While{iterator.hasNext()}{
      $s =$ iterator.next()\;
      newcandidate = \textbf{new} Set($S$)\;
      newcandidate.insert($s$)\;

      iterators.put(newcandidate,iterator.expand($s$))\; \label{algo:greedy:expand}

      \If{newcandidate.size()==$n$}{
        \If{result.size() $< k_g$}{
          result.insert(newcandidate)\;
        }\Else{
         \If{$\sigma_q$(newcandidate) $< \sigma_q$(result.peek())}{
            result.pop()\;
            result.insert(newcandidate)\;
          }\Else{
            \textbf{break}\;
          }
        }
      }\Else{
        counter = counter + 1\;
        \If{counter $\geq k_g$}{
          \textbf{break}\;
        }
      }
    }
  }
  \caption{greeverso ($q, n, R, k_g$): Greedy diversification algorithm.}
  \label{algo:greedy}
\end{algorithm}

In the heart of our paradigm is Algorithm \ref{algo:singlepoint}, which 
given a query center $q$ and a set of interconnected objects $S$, it returns 
the most diversified object $u$ in the network that also belongs to set $R$.
We note that the inclusion requirement for set $R$ aims at preventing 
certain web-pages from appearing in the result-set for reasons that do 
not concern this work, e.g., inappropriate or offensive content, reported 
web-pages, etc. Alternatively, one could remove those elements from the 
graph, but this is a more subtle and elegant intervention that can be upheld 
if necessary. Besides, all documents that were led from those nodes and 
were not accessible via any other path, would not be accessible anymore, 
regardless whether they belong to $R$ or not. Thereby, $R$ could play the role 
of a while-list or a black-list, depending on the requirements of the problem.

We first initialize the min-heaps used for searching around $q$ and each 
element of $S$ in lines 1--11, Alg.~\ref{algo:singlepoint}. More importantly, 
the vertices are ranked in terms of an aggregation of network distance and 
context similarity according to either Eqs.~\ref{eq:avg} and \ref{eq:max}. 
For each source vertex, we have a heap where the vertices are inserted according 
to their distance from that particular source, and another in terms of their 
partial score regarding that source only, though. By convention, we let the 
complete score of an individual vertex $u$ given by $\sigma_q(S\cup\{u\})-
\sigma_q(S)$, and it reflects the change in the score of the augmented set. 
For instance, it follows for Eq.~\ref{eq:avg} that the change incurred in 
the score of the diversified set is equal to 
\begin{equation}
  \frac{1}{|S|+1} (\lambda \alpha d^{graph}(q,u) 
  - \beta \frac{1-\lambda}{|S|} \sum_{s \in S} d^{graph}(s,u)) 
  + \frac{|S|}{|S|+1} \sigma_q(S) \nonumber
\end{equation}
in terms of its network properties, and in terms of its context is
\begin{equation}
  \frac{1}{|S|+1} (\lambda (1-\alpha) d^{text}(q,u) 
  - (1-\lambda) \frac{1-\beta}{|S|} \sum_{s \in S} d^{text}(s,u)) 
  + \frac{|S|}{|S|+1} \sigma_q(S) \nonumber
\end{equation}
Of course, since we want to retrieve the object with the smallest such value 
by comparing a carefully selected subset of the available web-documents with 
each other, we can just use the part of the formula within the parentheses and 
disregard the rest for it makes no difference when comparing any two web-pages.

Moreover, this marginal gain can be further divided accordingly in the partial 
costs arising from each of the elements of $S$ or the query center $q$. Likewise, 
we obtain the cost function associated with Eq.~\ref{eq:max} which has a branching 
form. In particular, we discern the following cases: 
(i) $d^{\text{rel}}(p,q) \leq 
\max_{x \in S} d^{\text{rel}}(x,q)$ and $\min_{x \in S} d^{\text{diss}}(p,x) \geq 
\min_{y,z \in S} d^{\text{diss}}(y,z)$, and thus, the marginal gain $\sigma_q(p)$ 
caused by the insertion of $p$ is equal to $0$, 
(ii) $d^{\text{rel}}(p,q) > \max_{x \in S} 
d^{\text{rel}}(x,q)$ and $\min_{x \in S} d^{\text{diss}}(p,x) \geq \min_{y,z \in S} 
d^{\text{diss}}(y,z)$, and hence, we have $\sigma_q(p) = \lambda (d^{\text{rel}}(p,q)-\max_{x \in S} 
d^{\text{rel}}(x,q))$, 
(iii) $d^{\text{rel}}(p,q) \leq \max_{x \in S} d^{\text{rel}}(x,q)$ 
and $\min_{x \in S} d^{\text{diss}}(p,x) < \min_{y,z \in S} d^{\text{diss}}(y,z)$, 
then it follows $\sigma_q(p) = (1-\lambda)(\min_{x,y \in S} d^{\text{diss}}(x,y) - \min_{z \in S} 
d^{\text{diss}}(p,z))$, 
(iv) otherwise, we take, $\sigma_q(p) = \lambda 
(d^{\text{rel}}(p,q)-\max_{x \in S} d^{\text{rel}}(x,q)) + (1-\lambda)(\min_{y,z \in S} 
d^{\text{diss}}(y,z) - \min_{x \in S} d^{\text{diss}}(p,x))$; 
where $d^{\text{rel}}(.,.)$ 
denotes the affine combination of network distance and content similarity for the relevance 
factor (smoothed using the $\alpha$ parameter), and $d^{\text{diss}}(.,.)$ stands for the 
affine combination that corresponds to the result items' dissimilarity (smoothed with the 
$\beta$ parameter) from Eq.~\ref{eq:max}.
Hence, between any two objects we are inclined to prefer the one whose insertion 
leads to the set of the better rank. 

\begin{algorithm}[!htbp]
  initialize score min-heap for $q$\;
  insert $q$ with $0$ score value\;

  initialize distance min-heap for $q$\;
  \ForEach{adjacent node of $q$}{
    insert it in terms of network distance from $q$\;
  }

  \ForEach{$s \textbf{ in } S$}{
    initialize score min-heap for $s$\;
    insert $s$ with $0$ score value\;

    initialize distance min-heap for $s$\;
    \ForEach{adjacent node of $s$}{
      insert it in terms of network distance from $s$\;
    }
  }

  create candidates min-heap congregating the tentative results\;

  \While{the best element of the candidates is no better than the score of the best undiscovered node}{
    $h_\sigma$ the min cost expansion score heap\;
    $h_\delta$ the respective distance heap for the same source, say $u$\;

    $w$ = $h_\sigma$.pop()\;

    \If{$w$ discovered from all sources}{
      \If{$w$ \textbf{in} $R$}{
        push $w$ to the candidates\;
      }
    }\Else{
      update partial score of $w$, since now accessed from $u$\;
    }

    \While{$\sigma_q$ ($h_\sigma$.peek()) $\geq \gamma \times d^\text{graph}(u,h_\delta$.peek())}{

      $v$ = $h_\delta$.pop()\;
      $h_\sigma$.push($u$)\;

      \ForEach{adjacent node $y$ of $v$}{
        \If{$y$ undiscovered from $u$}{
          $h_\delta$.push ($y$,$v$.weight+$(v,y)$.weight)\;
        }
      }
    }
  }
  \Return candidates.pop()\;
  \caption{verso ($q,S,R$): Graph-context diversification given a query vertex $q$, 
                          its context and a set of vertices $S$ with their contexts.}
  \label{algo:singlepoint}
\end{algorithm}

Next, in lines 13--27, we search for the most diversified vertex by expanding with each 
iteration our search from the source node that contributes the least partial score among 
$q$ and all elements of $S$. If the newly accessed node $u$ is now visited from all source 
nodes then it is most certainly qualified to be included in the set of the candidate solutions, 
provided that it also belongs to set $R$. Otherwise, we need to update the partial score of 
$u$ accordingly. Furthermore, since the sources from which we initiate our graph search may 
lie distant from each other, the structure where we keep the vertices by their partial scores 
may grow disproportionately large until we retrieve the very first vertex that is accessed from 
all sources. This situation can turn out extremely tricky when the sources from which we initiate 
the search are far away from each other in terms of their network distance, and this case is not 
necessarily uncommon in practice. Therefore, it is of major importance to store this information 
regarding vertices' partial scores in such a way that it can be accessed and updated effectively 
and efficiently. In particular, we need to check in line 13 whether the best score that any vertex, 
which is not yet accessed from all sources, is outranked by the best candidate solution (and thus 
encountered from all sources) retrieved so far, and we are therefore in position of returning it 
to the user. If this is not the case, the search needs to be continued with more iterations until 
the constraint is satisfied. Nevertheless, in order to compute a meaningful lower bound for the 
best possible complete score of any partially accessed vertex, we will have to put at the place 
of the missing information (after all we are talking about partial scores) the distance value that 
corresponds to the reach of the search from the source node that the examined node has not been 
visited yet. Hence, even if we have those nodes sorted according to their partial scores, we still 
have to enrich that information with the respective distances, so as to safely compute the termination 
criteria without any loss. Most importantly, this information changes constantly with every iteration 
of the while-loop in lines 13--27. Thereby, the most appropriate structure to accommodate this operation 
along with fast and efficient updates is a sorted list, where each time we find the position of each 
element to be inserted with binary search. Hence, by selecting the first $C$ elements with the minimum 
partial scores at the head of the list, where $C$ takes a small value like $3$ and $5$, and replacing 
the respective distances when appropriate, we find the best possible score value that any partially 
accessed vertex could achieve. But now there might be partial scores that fall below that value. Hence, 
we need to examine all the elements of the list that have a partial score below that value, and among 
those values we shall find the best possible score an unvisited vertex can achieve. 

Similarly, with each iteration of the while-loop, we need to update in line 
21 the partial score of the newly accessed node since it is now accessed from 
a new source. Next, the new position of the accessed node is found using binary 
search. Using a different data-structure for performing these operations, for 
example a heap, would involve extremely expensive updates. Similarly, examining 
linearly all the vertices would also be computationally expensive since the 
size of the structure could grow very large to perform these operations again 
and again, as with each iteration we have a new distance that would complement 
a portion of the scores of the partly seen nodes differently than before. In 
line 16, we pop the element that contributes the minimum score increase (low 
score values correspond to a better rank) among the heads of all score-heaps, 
and if that node is now accessed from all sources, it will be inserted in the 
heap where we keep all candidate results ordered by their scores; otherwise 
we update its partial score accordingly in line 21. Then, in lines 22--27, 
we pop from the distance-heap that is associated with the same source the 
next closest vertex and we insert it into the respective score-heap, for 
this is the very next node, among the ones not encountered from all sources, 
to cause the least score overhead given the Maximal Marginal Relevance (MMR) 
ranking functions that we study in this paper. Then, in lines 25--27, we 
augment the respective distance-heaps with all the adjacent nodes of the 
popped element that are unvisited from that specific source, so as to be 
able in the next iteration of the while-loop to access the next closest 
vertex. The purpose of this nested inner loop is to ensure that each time 
we pop an element from a score heap that is associated with a specific 
source, there is no other vertex in the network that can cause a smaller 
increase to the score function given that source. This is the case where 
the constraint of the while-loop is satisfied in line 22, as it is ensured 
that the next shortest distance multiplied by either factor $\alpha$ or 
$\beta$, depending on whether the source that we examine at the time is 
the query center or any of the elements of $S$, is greater than the least 
score value in the respective heap. Hence no textual description considered 
here, outranks the best element at the head of the associated score heap. 
The combination of the two loops ensures that each time we return the vertex 
that achieves the optimal score overall, given that specific query center 
$q$ and the set of vertices $S$ that were given as arguments.

Alg.~\ref{algo:singlepoint} is used to implement the iterator of our aforementioned methods 
Algs.~\ref{algo:interchange} and \ref{algo:greedy}. Remember the GraphDivIterable object that 
we create in line \ref{algo:interchange:iterator}, Alg.~\ref{algo:interchange} and line 
\ref{algo:greedy:iterator}, Alg.~\ref{algo:greedy}. In particular, lines 1--12 would be part 
of the constructor of the iterator while the lines 13--28 would constitute the method returning 
the next best object in the graph to complement the set formed so far. A function testing whether 
there are still objects to be retrieved would simply check whether there are still objects in the 
candidates' heap or any of the score-heaps and distance-heaps with items waiting to be processed. 
Moreover, in line \ref{algo:interchange:replace} in Alg.~\ref{algo:interchange} and line 
\ref{algo:greedy:expand} in Alg.~\ref{algo:greedy} we use the previous state of the existing 
iterators to enhance the newly created ones and prevent from repeating a significant portion of 
the graph search. This is an important optimization since graph traversal might consume considerable 
time. In addition, when we create a new iterator, we insert into its list with the partially seen 
objects all candidate objects from the iterator we copy the state to initialize the new one. Those 
items are now seen from all sources but the new source for the new iterator. In addition, we copy 
the elements from the old list with partially seen objects, and also, the objects that had already 
been returned as results (we keep these in a list).

Further, due to the nature of the problem, when real-time response is required 
for processing very large graphs, we can set a time-out parameter for each of 
the two modules of our implementation for the greedy and the hill-climbing methods. 
Equivalently, we can limit the maximum number of iterations by adding an additional 
such constraint in line 16, Alg.~\ref{algo:interchange} and line 12, Alg.~\ref{algo:singlepoint}. 
This parameter would have no meaning for small sets, but might be required for 
large diversified sets in order to achieve faster response times at the cost of 
a result of lesser quality though. First, the most diversified object in the network 
is returned within $t_d$ time, the time-out parameter we could add in line 12, 
Alg.~\ref{algo:singlepoint}. When the execution of the loop is stopped due to the 
time-out variable, most probably the algorithm has not returned the next most 
diversified object, but instead, the best candidate that has been retrieved so 
far. This is an indirect way since setting a cut-off time parameter for our 
greedy method in Alg.~\ref{algo:greedy} as a whole is prohibited since sets with 
less than $n$ elements could be returned otherwise. On the other hand, for our 
hill-climbing based algorithm, we set cut-off time $t_c$ to a multiple of $|S| 
\times k_g \times t_d$, so as to allow for the refinement of all the seeds by at 
least a few elements. Nevertheless, when a fast response is required and $t_c$ 
takes a small value, parameter $k_c$ should also be quite small, and therewith, 
just one or two of the best seeds will be further processed. Whether all of those 
selected seeds will be refined further also depends on the time-out parameter, 
and thus, for low $t_c$ values setting $k_c$ appropriately small would allow for 
those few seeds to reach their respective optima within the required time-frame. 
Under these circumstances, this is a wiser policy compared to expanding more seeds 
(for large $k_c$ values) inadequately.

\vspace{-5pt}
\section{Experimental Evaluation}
\label{sec:exp}

In this section we assess the performance of our methods for various 
configurations and demonstrate our results following.

\subsection{Setting}

\textbf{Metrics.} 
We adopt three important metrics: (i) the total \emph{execution 
time} required until each method returns its result, (ii) the allocated 
\emph{memory} by all data-structures (lists, queues, heaps, etc.), without 
considering the memory required for storing the respective graphs and 
web-documents' tf-idfs for the employed vector space model, (iii) the 
\emph{quality} of the result returned by each method. The scores of the 
respective cost functions are used for this purpose.

\textbf{Parameters.} 
A variety of different parameters have been employed in the 
construction of our synthetic workload so as to ascertain 
the effectiveness and efficiency of our paradigm: (1) the 
number of web-documents, (2) the number of links per page, 
(3) number of lemmas per web-document whose frequency follows 
a zipfian distribution, (4) the number of web-pages in the 
diversified set that is returned to the user, (5) the number 
of seeds $k_g$ which is also equal to $k_c$ in our configuration, 
(6) the compromise $\lambda$ between relevance and result 
elements' dissimilarity, (7) the network-context trade-off 
$\beta$ upon which result items' dissimilarity is computed. 
For (6) high values indicate a bias towards relevance, whereas 
in (7) high values indicate a bias towards graph characteristics 
over content. On the other hand, the relevance network/context 
trade-off is constant in our setting and considers documents' 
content exclusively as we fix $\alpha$ to $0$. Clearly, we are 
interested in a diversified result of highly relevant web-documents 
in terms of their content that are also diversified with each 
other in terms of both their content and network properties, 
a mixture smoothed by parameter $\beta$. In Tables 
\ref{table:dataset} and \ref{table:queryset}, we present all 
parameters along with the range of values that they take and 
their default values.

\begin{table}[hbt]
\centering
\begin{tabular}{c c c}
\hline
\textbf{Parameter}  & \textbf{Range}         & \textbf{Default} \\
\hline
\#web-pages         & $5k, 10k, 20k, 40k$    & $10k$\\
\#links per page    & $5, 10, 20, 40$        & $10$ \\
\#lemmas per page   & $50, 100, 200, 400$   & $100$\\
\hline 
\end{tabular}
\caption{Synthetic dataset parameters.}
\label{table:dataset}
\end{table}

\textbf{Datasets.} 
We make use of Wikipedia snapshots for real datasets 
that were obtained from \url{http://dumps.wikimedia.org/}.
The snapshot that we use here corresponds to the Greek 
collection of articles. It consists of 104,364 documents 
and in total there are 2,733,279 links. On average, each 
article has 17.47 links to other articles, while 16.7 
other links point to each article. From this collection 
we consider in the directed graph only the documents that 
correspond to actual articles and disregard user profiles, 
discussions, etc. For the synthetic graphs, we connect the 
vertices at random until we have the desired number of links.
For each configuration we execute 100 queries with query 
centers that are selected at random and arbitrarily. 

\begin{table}[hbt]
\centering
\begin{tabular}{c c c}
\hline
\textbf{Parameter}    &\textbf{Range}&\textbf{Default}\\
\hline
\#results             & $5,10,20,40$  & $10$ \\
\#seeds               & $1,2,3,4$     & $2$  \\
rel/diss $\lambda$    & $.2,.4,.6,.8$ & $.8$ \\
diss net/text $\beta$ & $.2,.4,.6,.8$ & $.8$ \\
\hline 
\end{tabular}
\caption{Queryset configuration.}
\label{table:queryset}
\end{table}

\begin{figure*}[!htbp]
  \centering
  \includegraphics[width=.1391\textwidth]{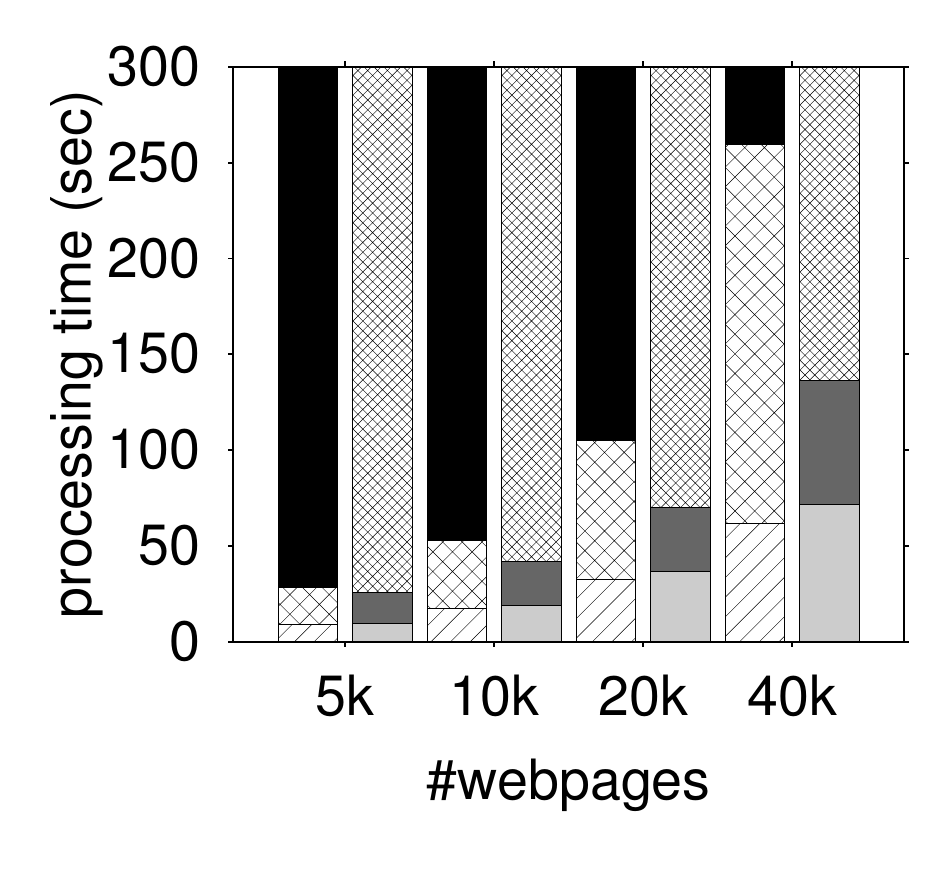}
  \includegraphics[width=.1391\textwidth]{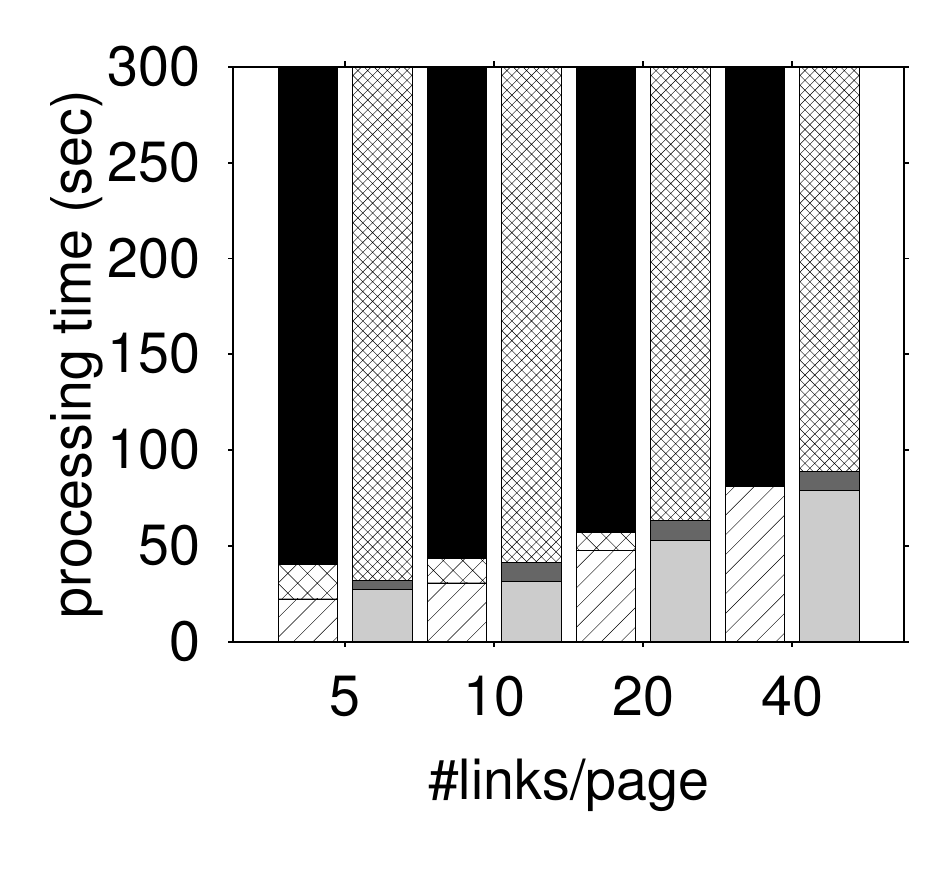}
  \includegraphics[width=.1391\textwidth]{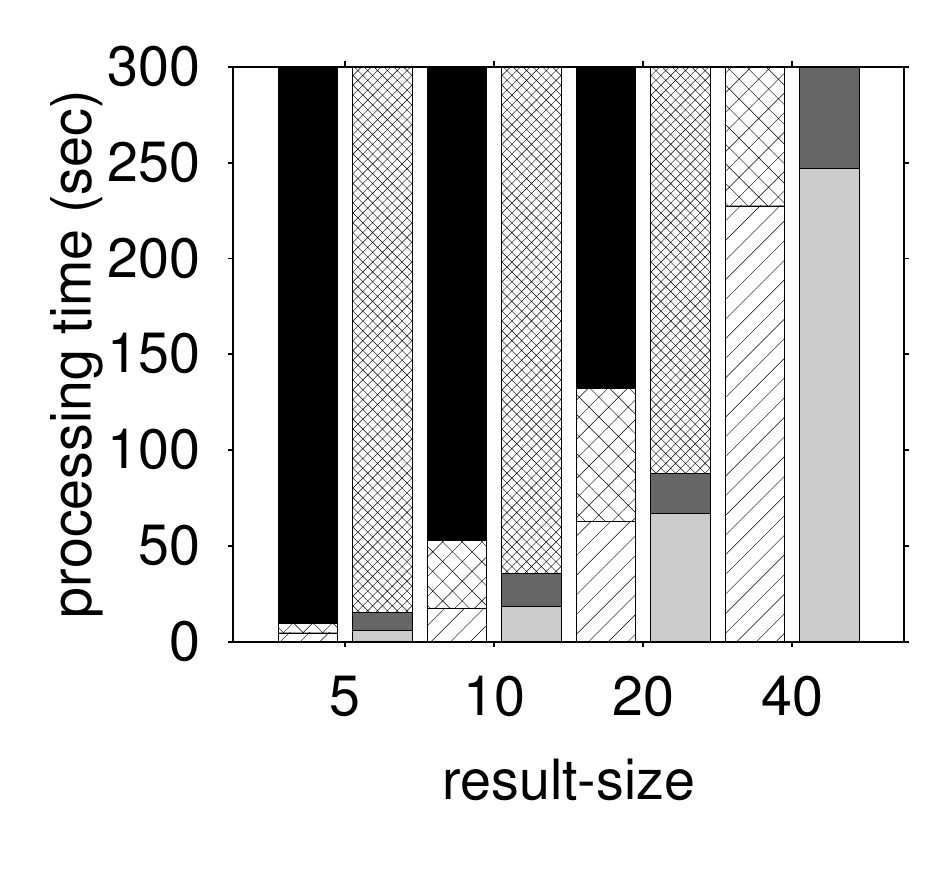}
  \includegraphics[width=.1391\textwidth]{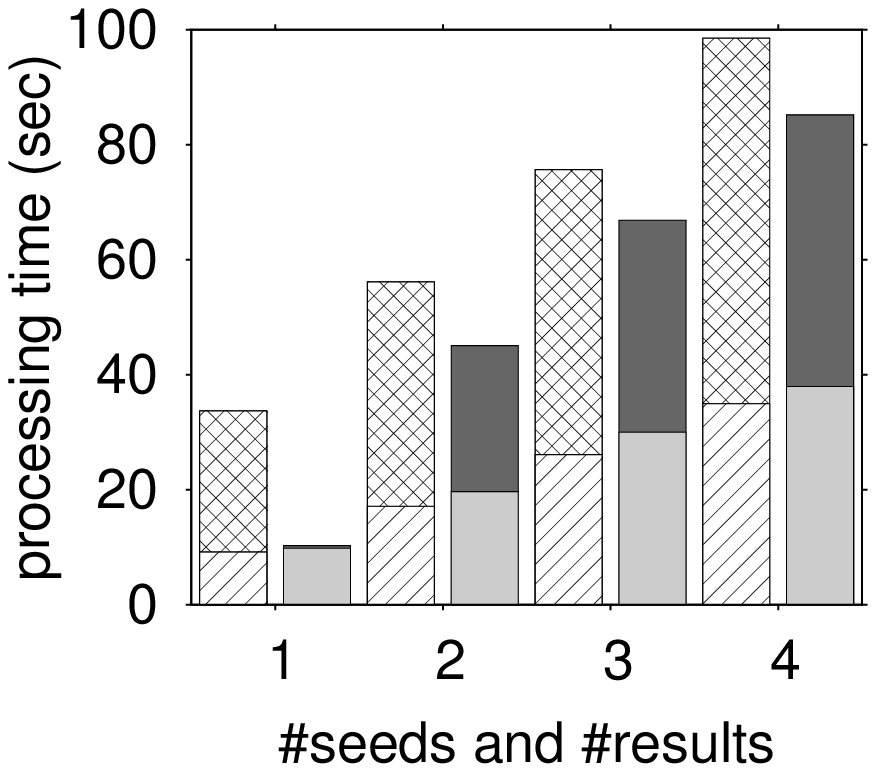}
  \includegraphics[width=.1391\textwidth]{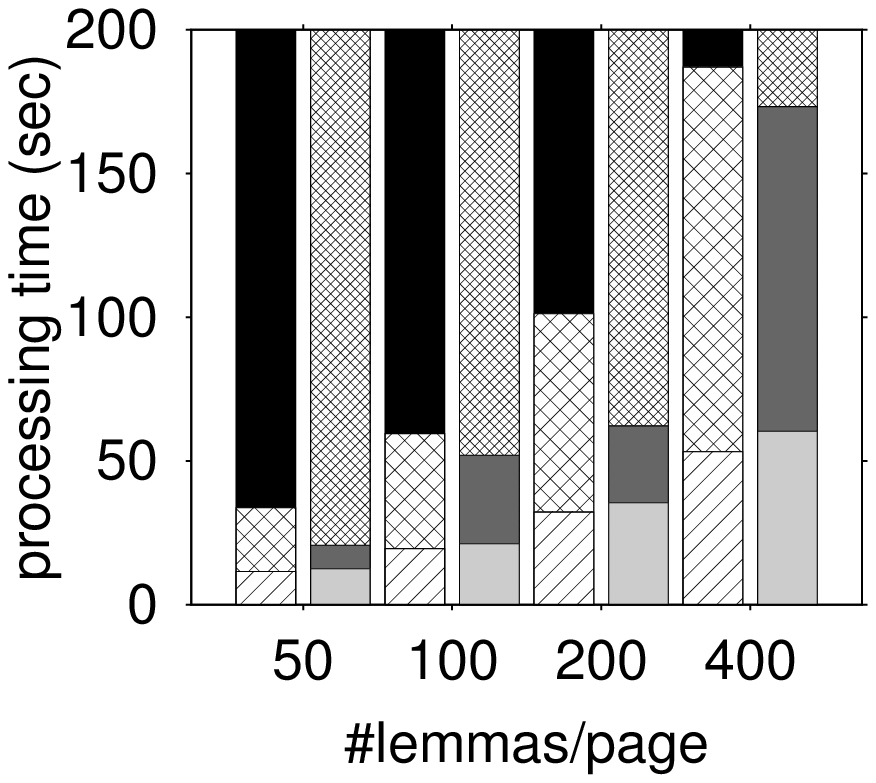}
  \includegraphics[width=.1391\textwidth]{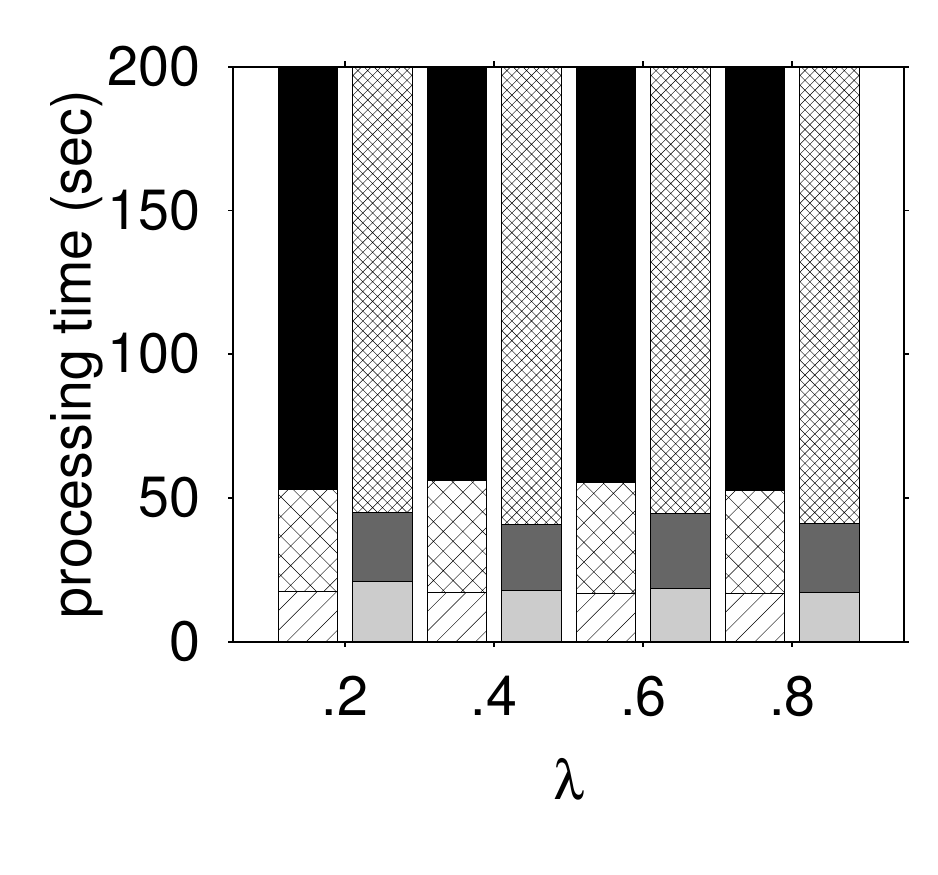}
  \includegraphics[width=.1391\textwidth]{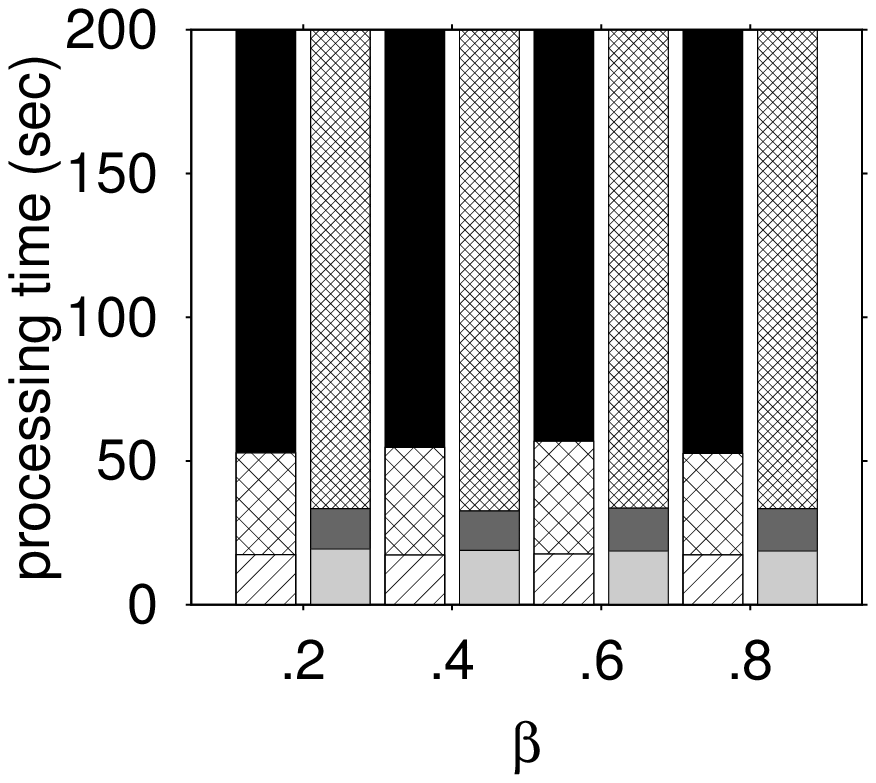}
  \vspace{-20pt}
  \caption{Processing time for synthetic workloads over various configurations.}
  \label{fig:synth:time}

  \centering
  \includegraphics[width=.1391\textwidth]{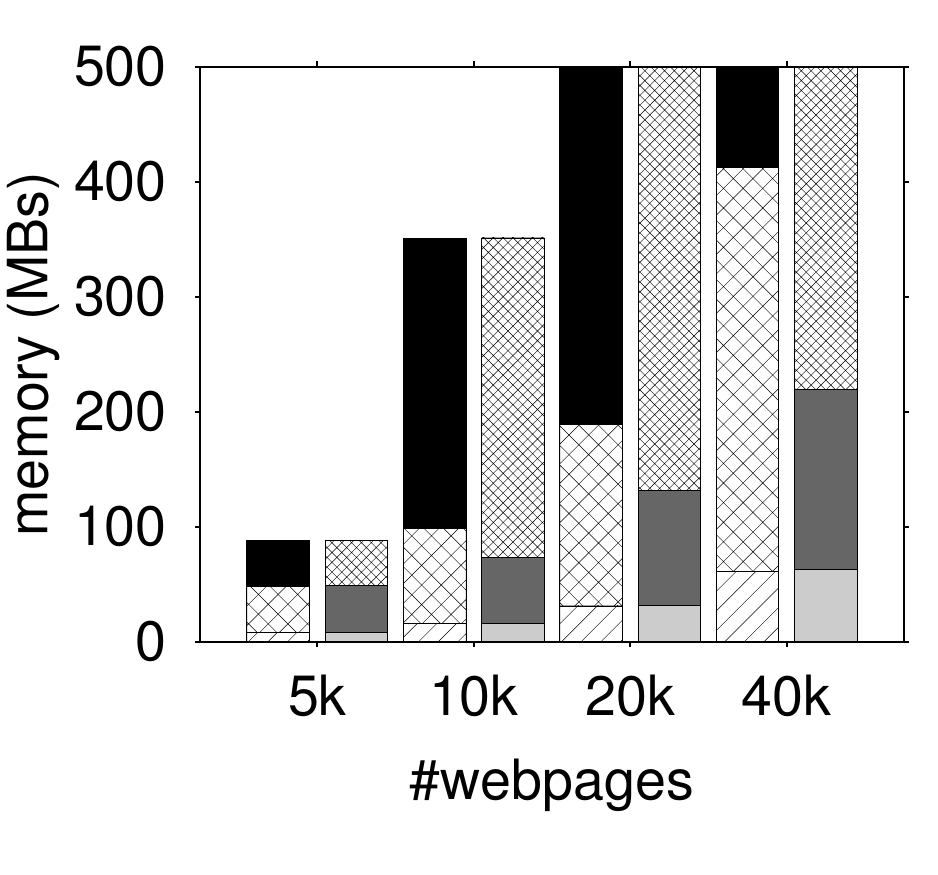}
  \includegraphics[width=.1391\textwidth]{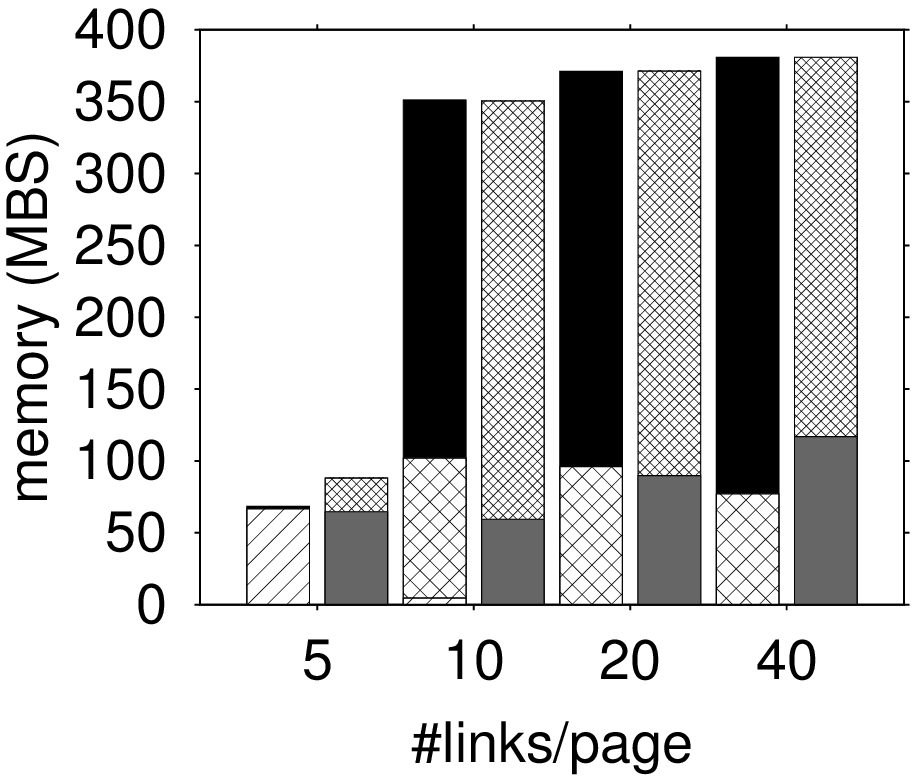}
  \includegraphics[width=.1391\textwidth]{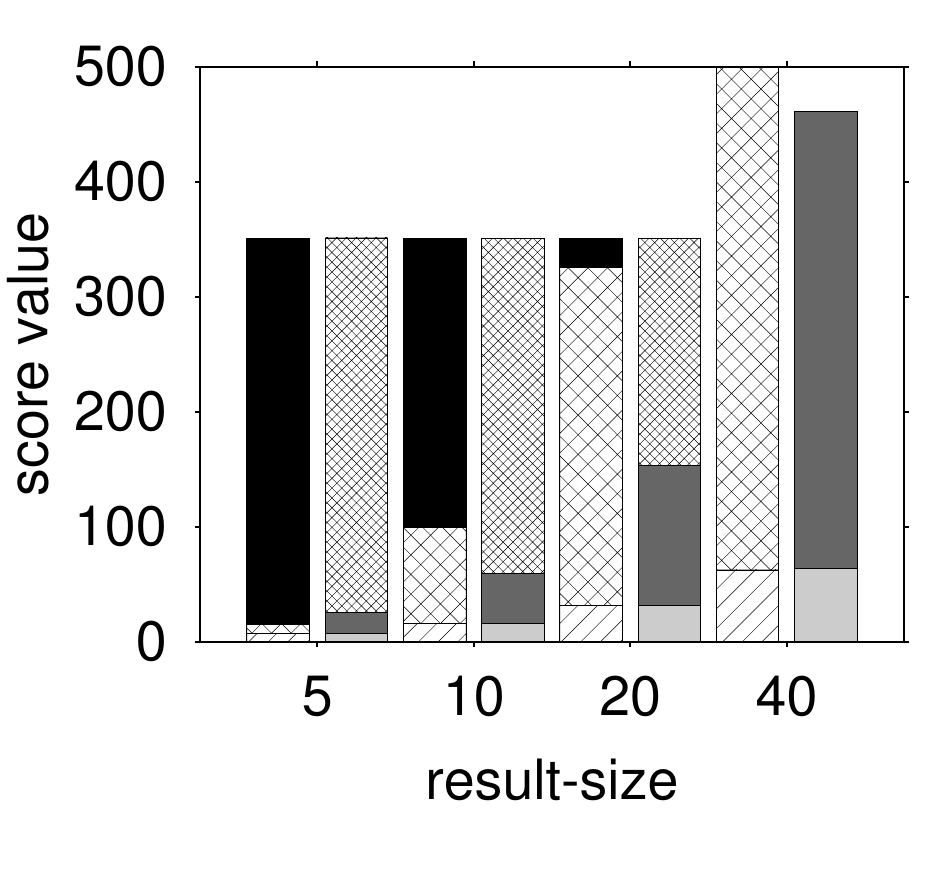}
  \includegraphics[width=.1391\textwidth]{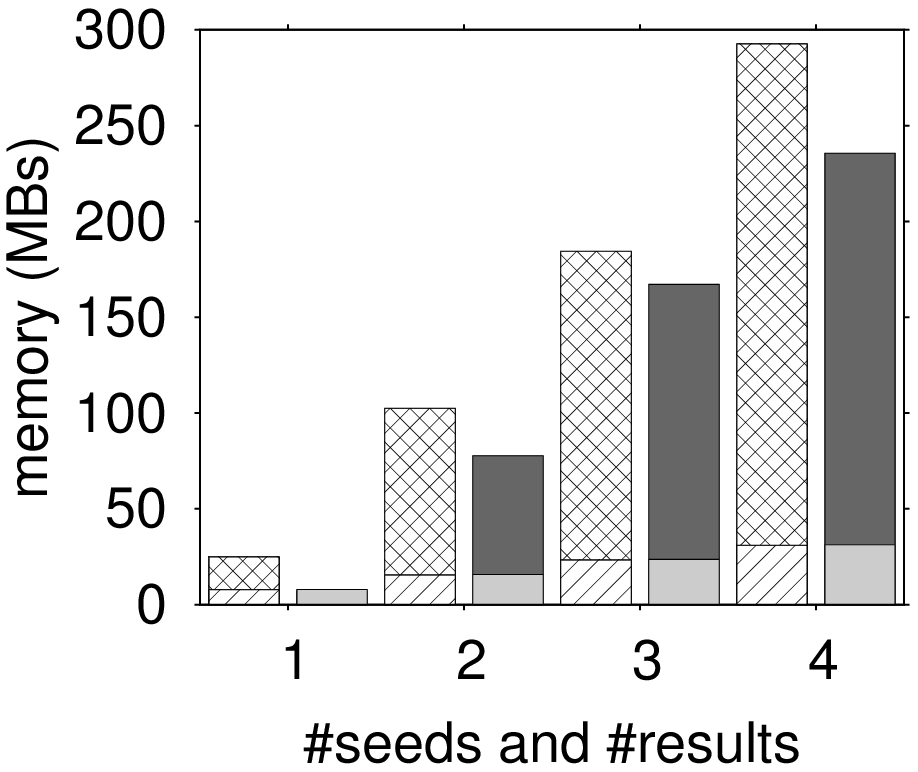}
  \includegraphics[width=.1391\textwidth]{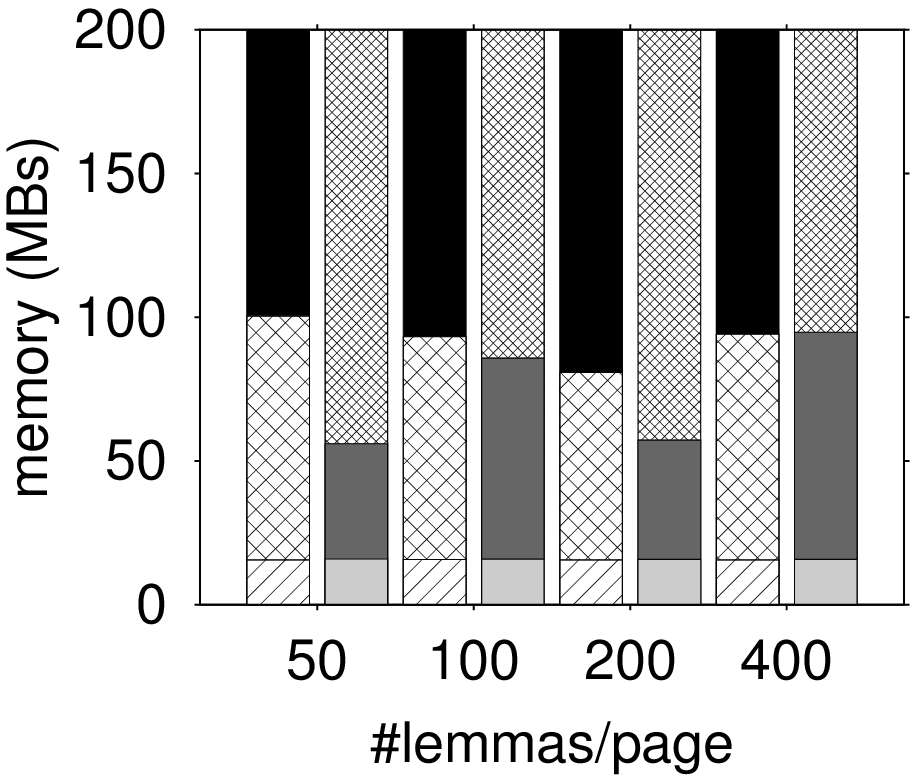}
  \includegraphics[width=.1391\textwidth]{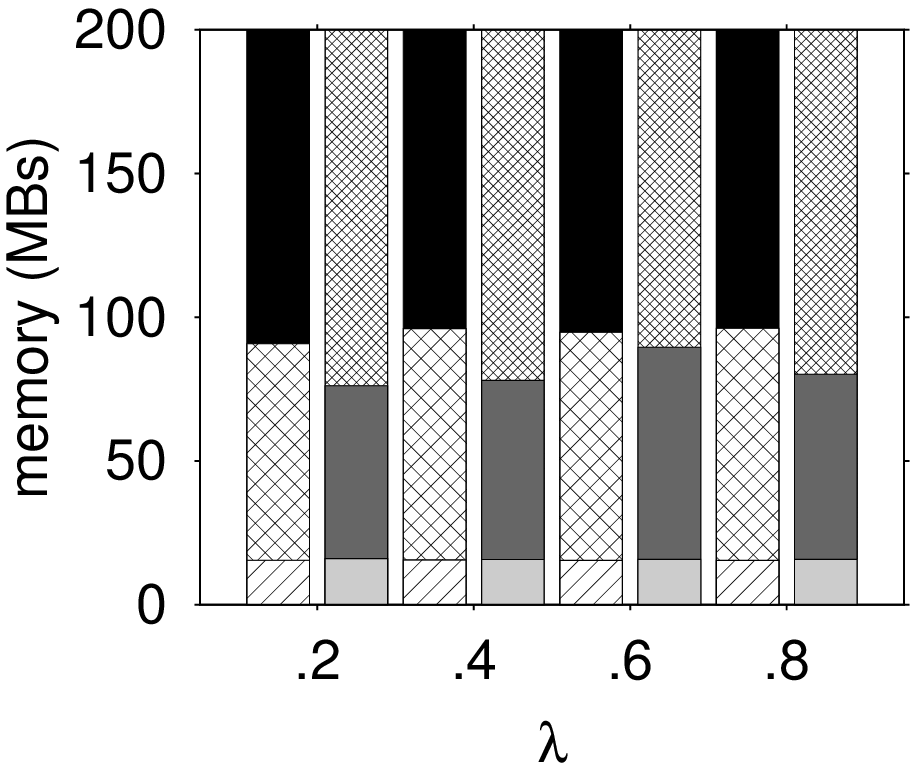}
  \includegraphics[width=.1391\textwidth]{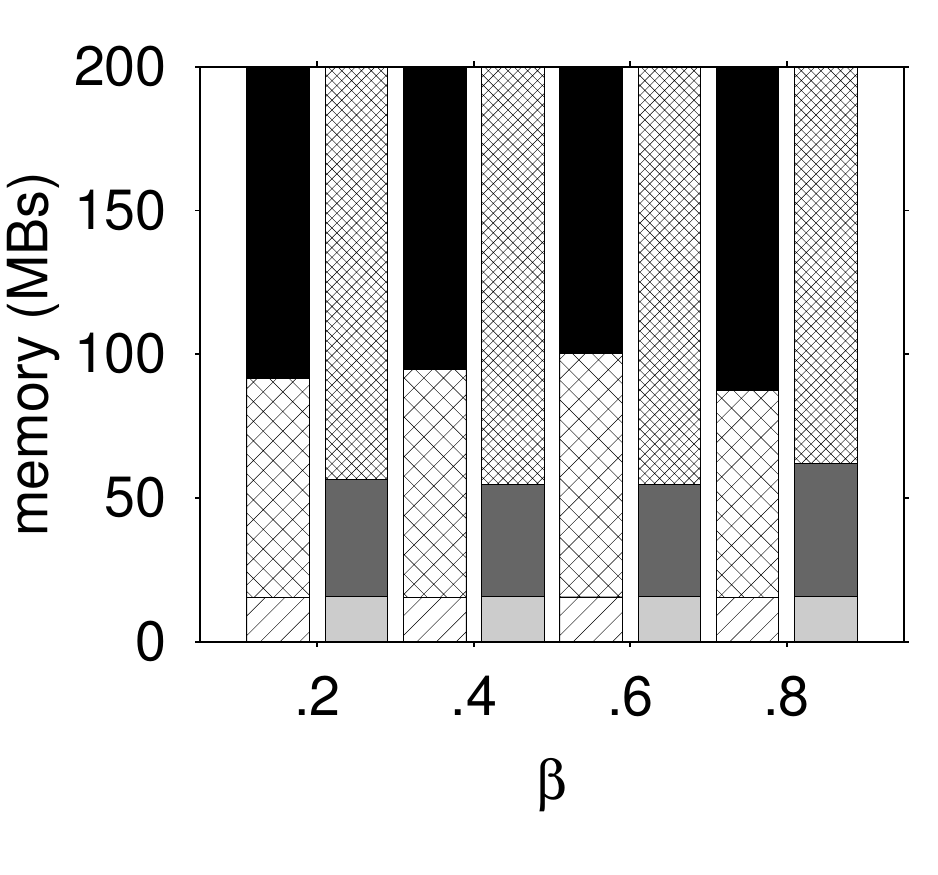}
  \vspace{-20pt}
  \caption{Allocated memory for synthetic workloads over various configurations.}
  \label{fig:synth:mem}

  \centering
  \includegraphics[width=.1391\textwidth]{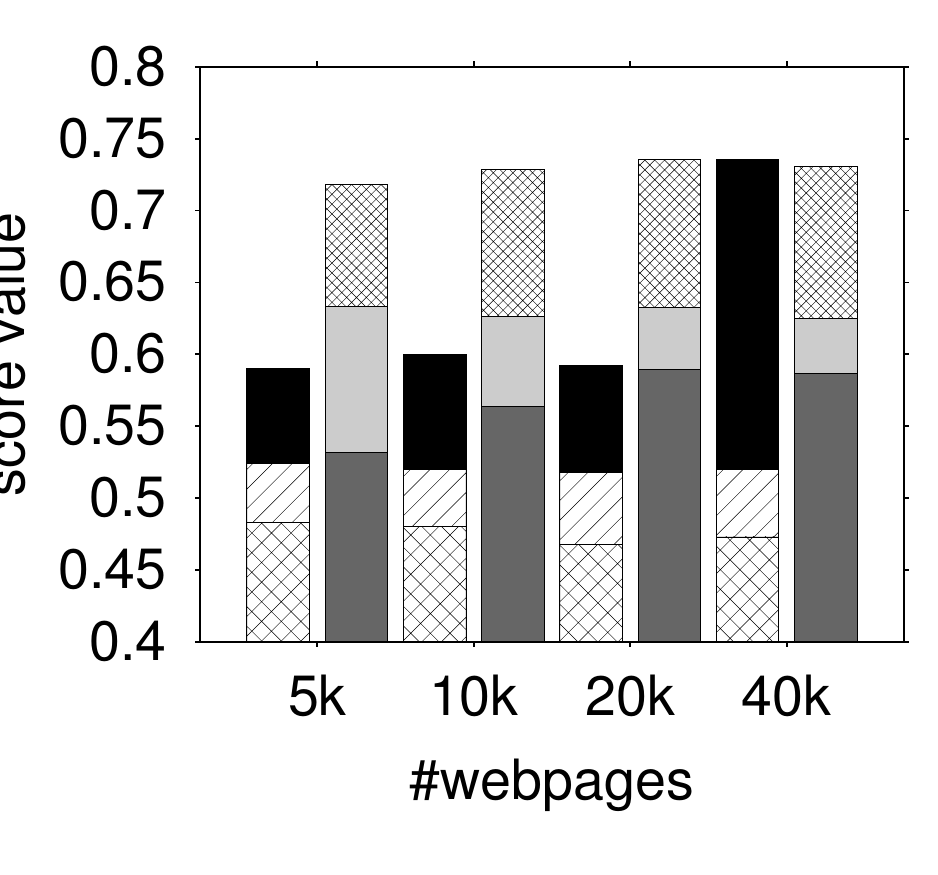}
  \includegraphics[width=.1391\textwidth]{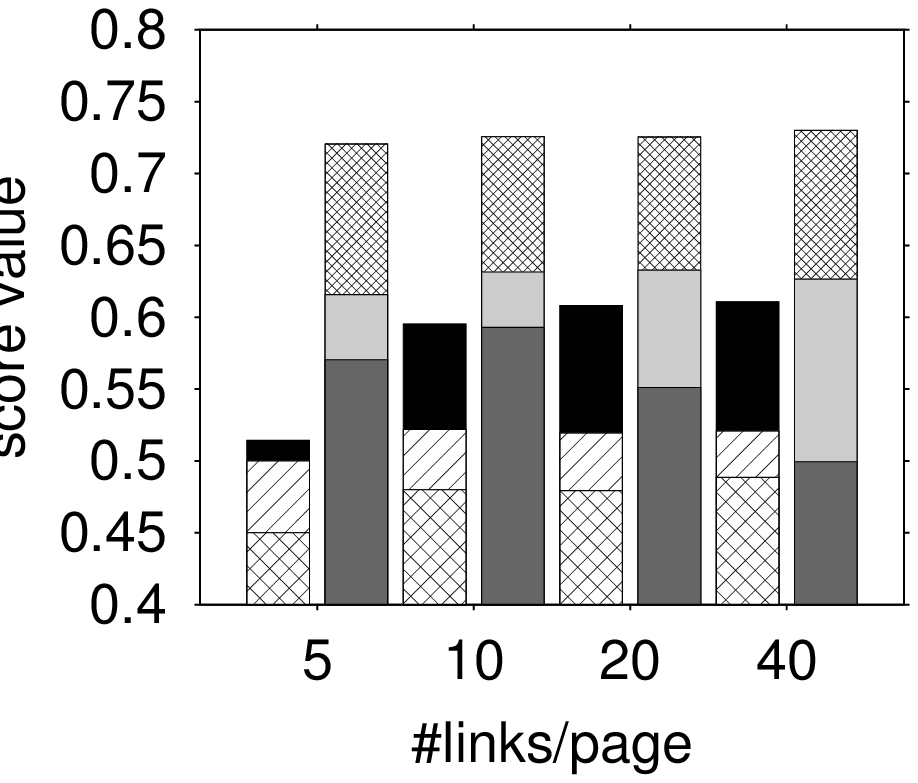}
  \includegraphics[width=.1391\textwidth]{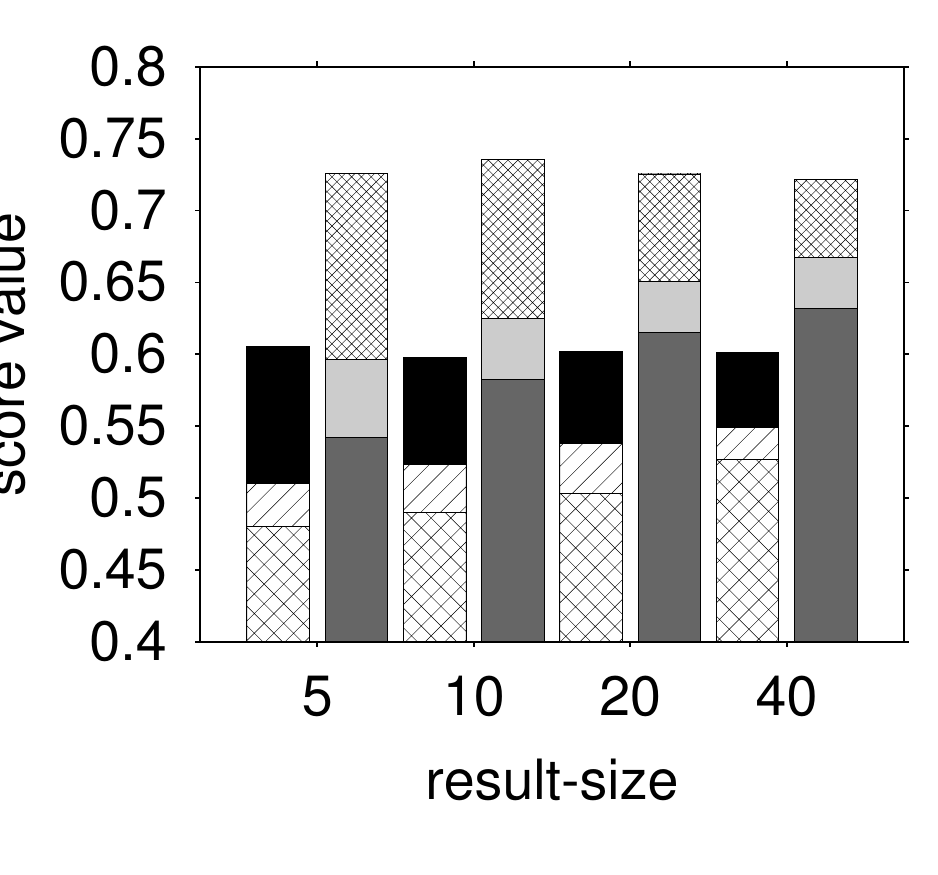}
  \includegraphics[width=.1391\textwidth]{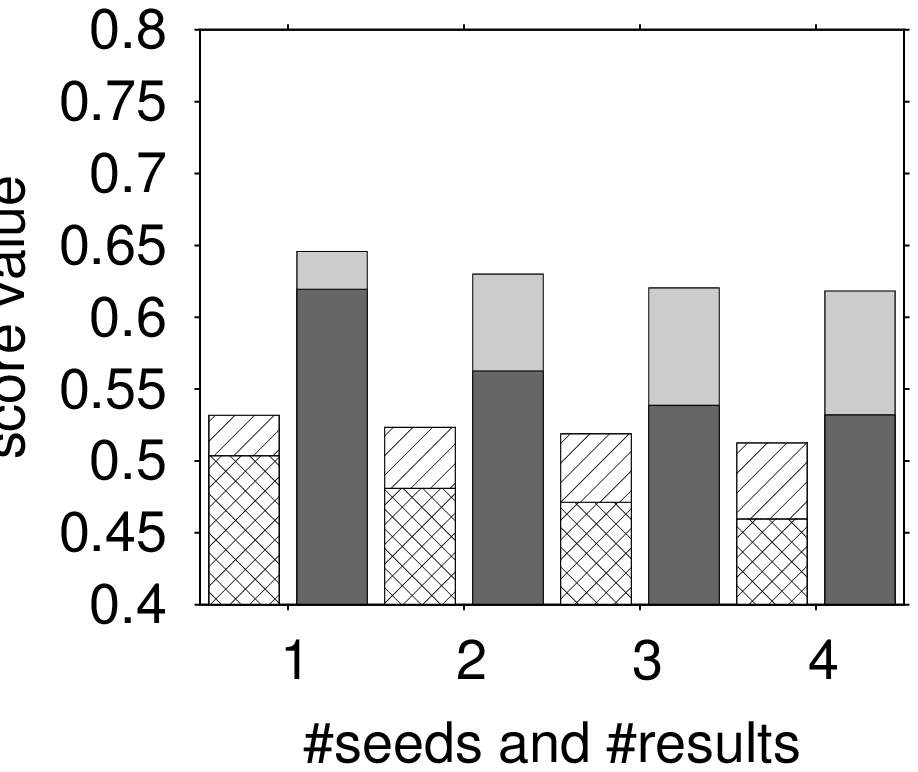}
  \includegraphics[width=.1391\textwidth]{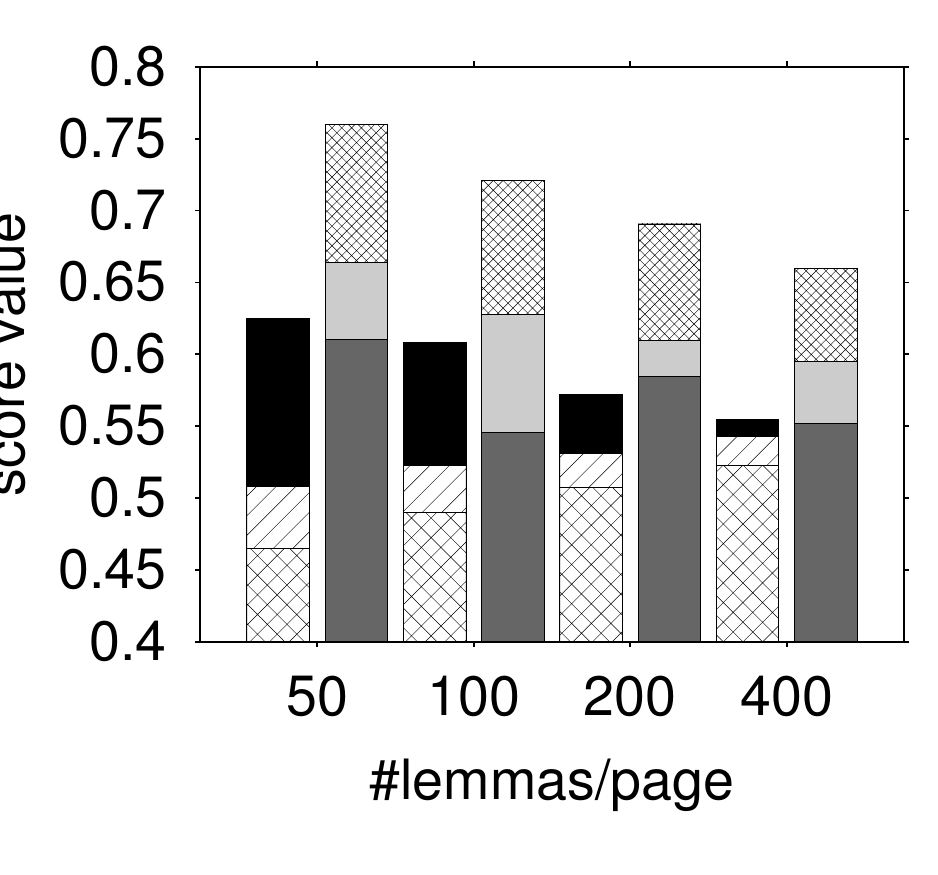}
  \includegraphics[width=.1391\textwidth]{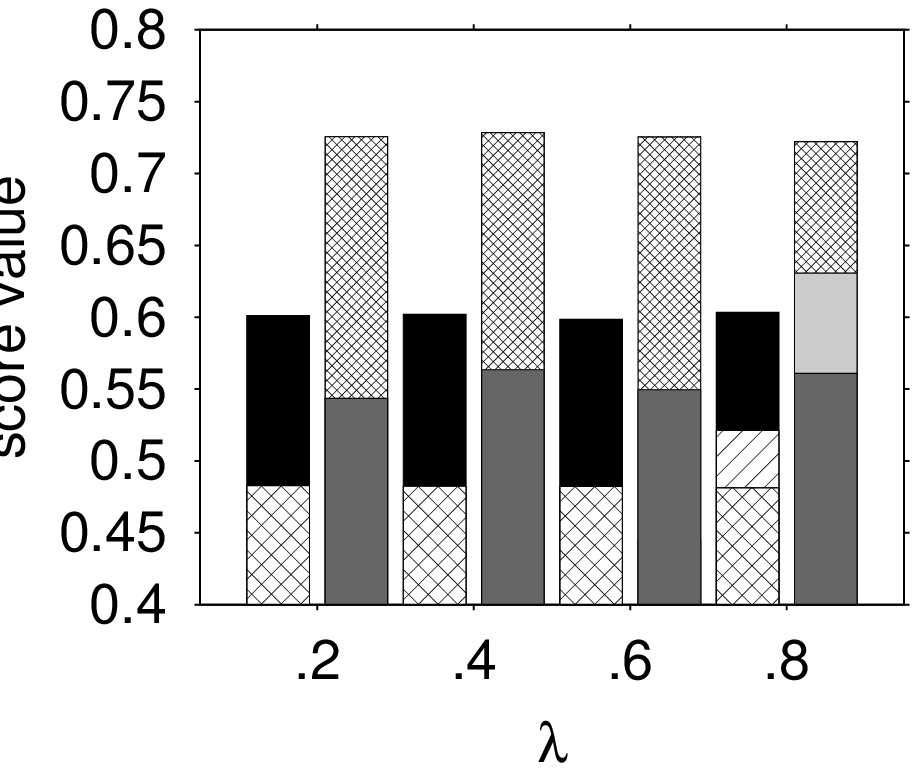}
  \includegraphics[width=.1391\textwidth]{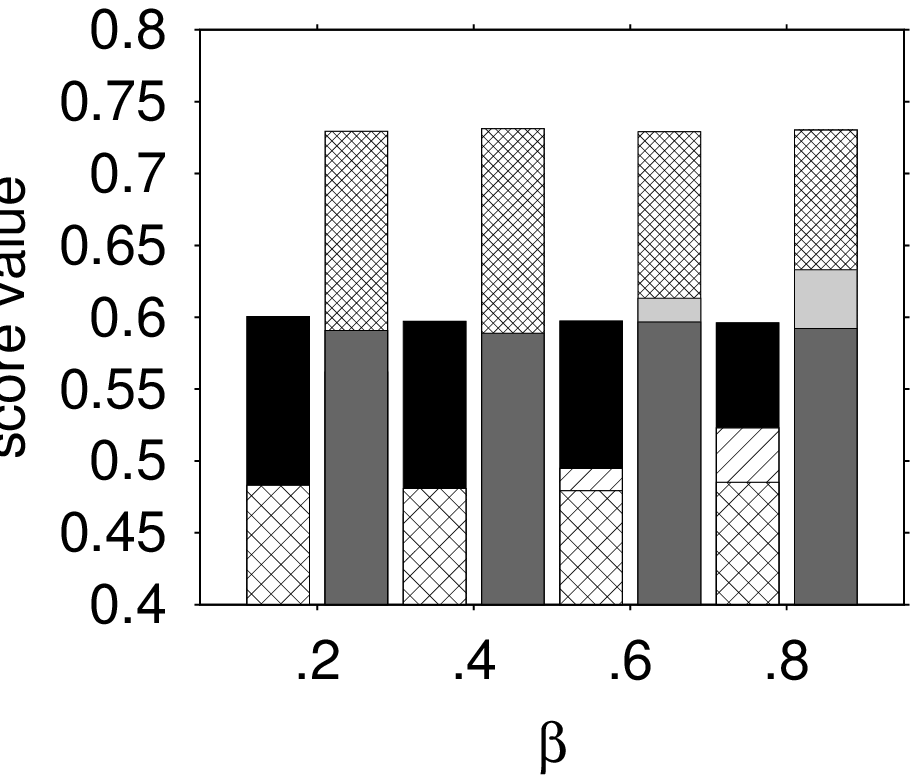}
  \vspace{-20pt}
  \caption{Rank function scores for synthetic workloads over various configurations.}
  \label{fig:synth:score}
\end{figure*}

\textbf{Baseline.} 
We compare our methods against an adapted version of 
\texttt{BestCoverage} from \cite{KucuktuncSKC13} for diversified 
ranking over graph data so as to consider documents' content as 
well. As we described in \S 2.2, their effort constitutes 
a heuristic scheme. All methods were implemented in Java and run 
in an AMD Opteron processor tweaked at 2.3 GHz.

\subsection{Results}

\begin{figure*}[hbt]
  \centering
  \includegraphics[width=.18\textwidth]{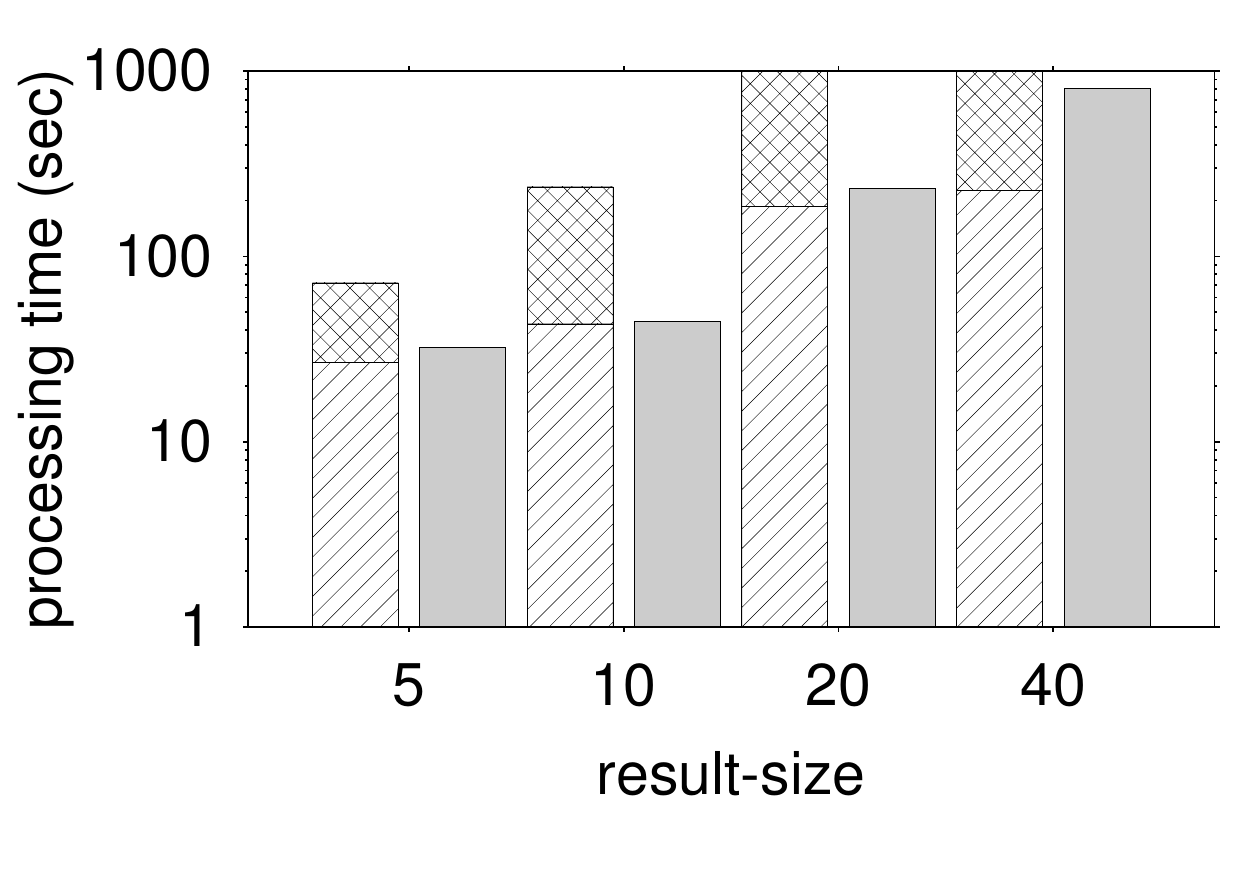}
  \includegraphics[width=.18\textwidth]{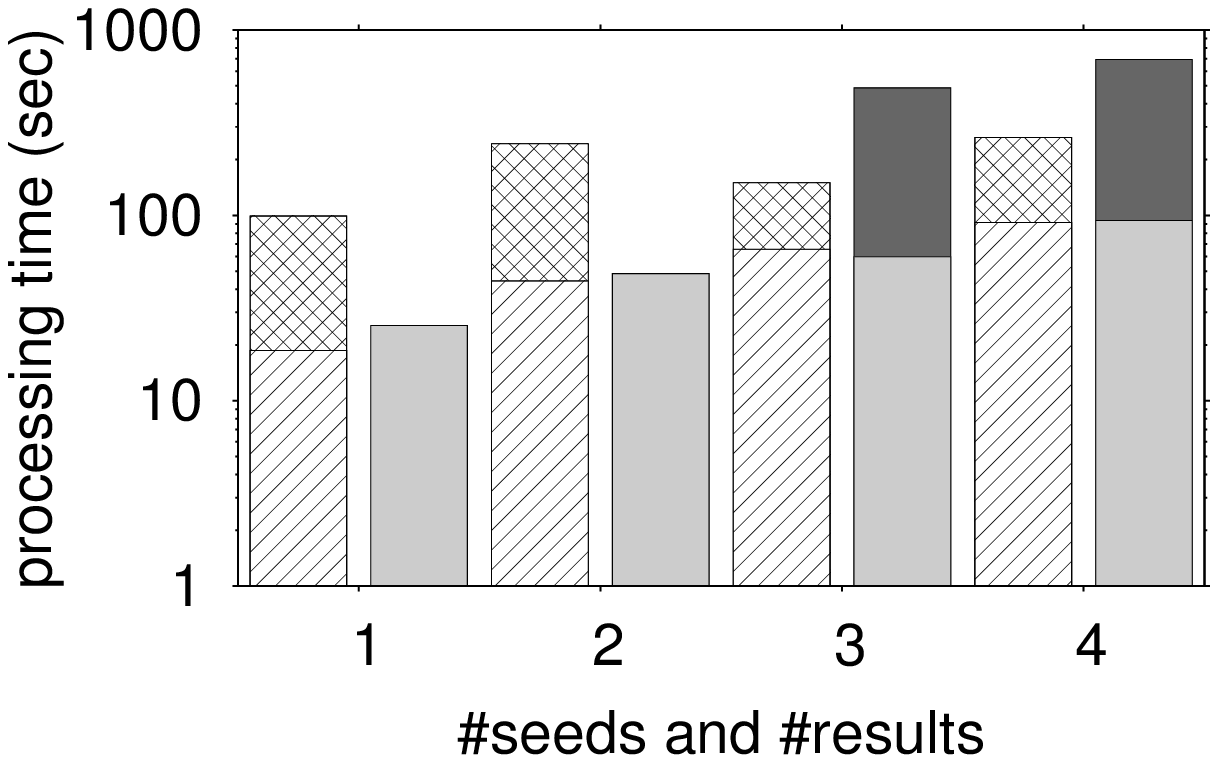}
  \includegraphics[width=.18\textwidth]{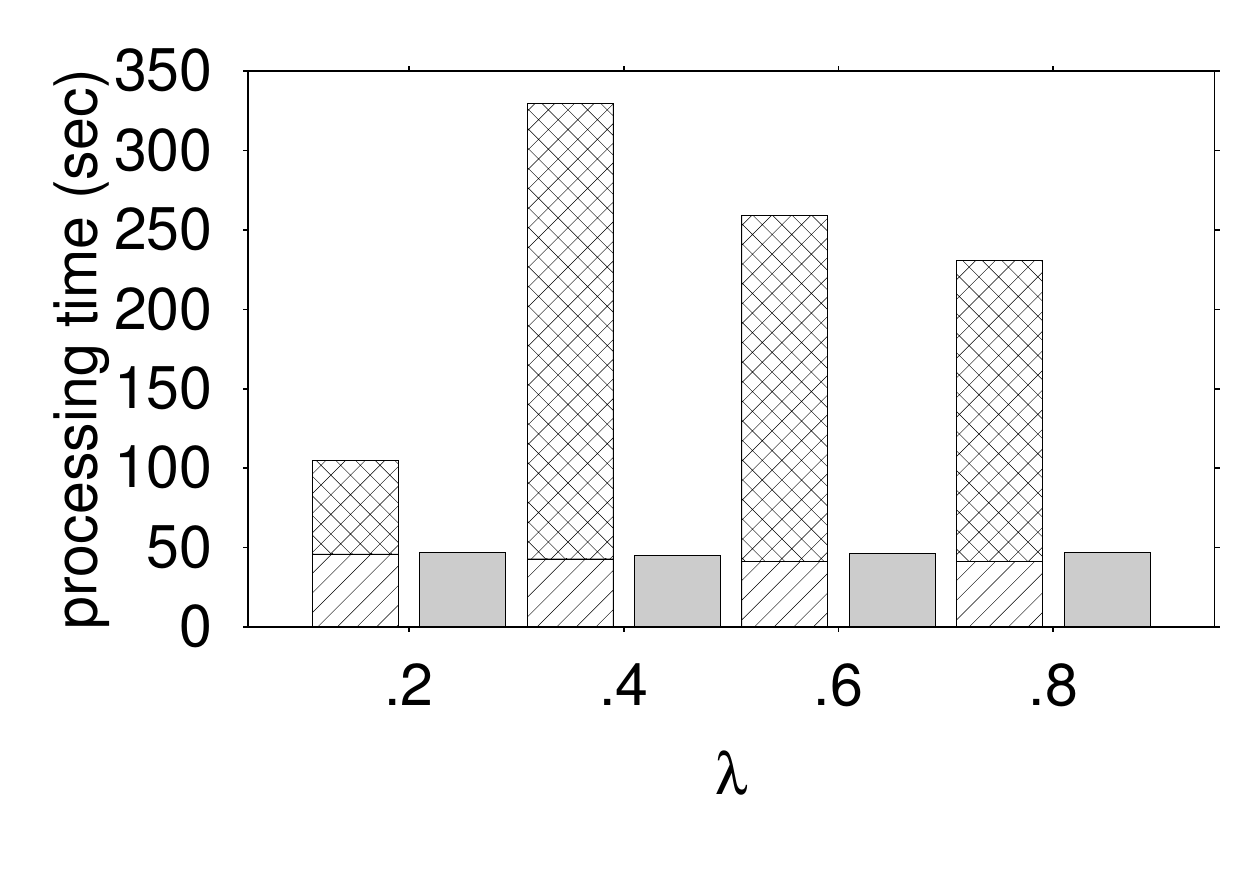}
  \includegraphics[width=.18\textwidth]{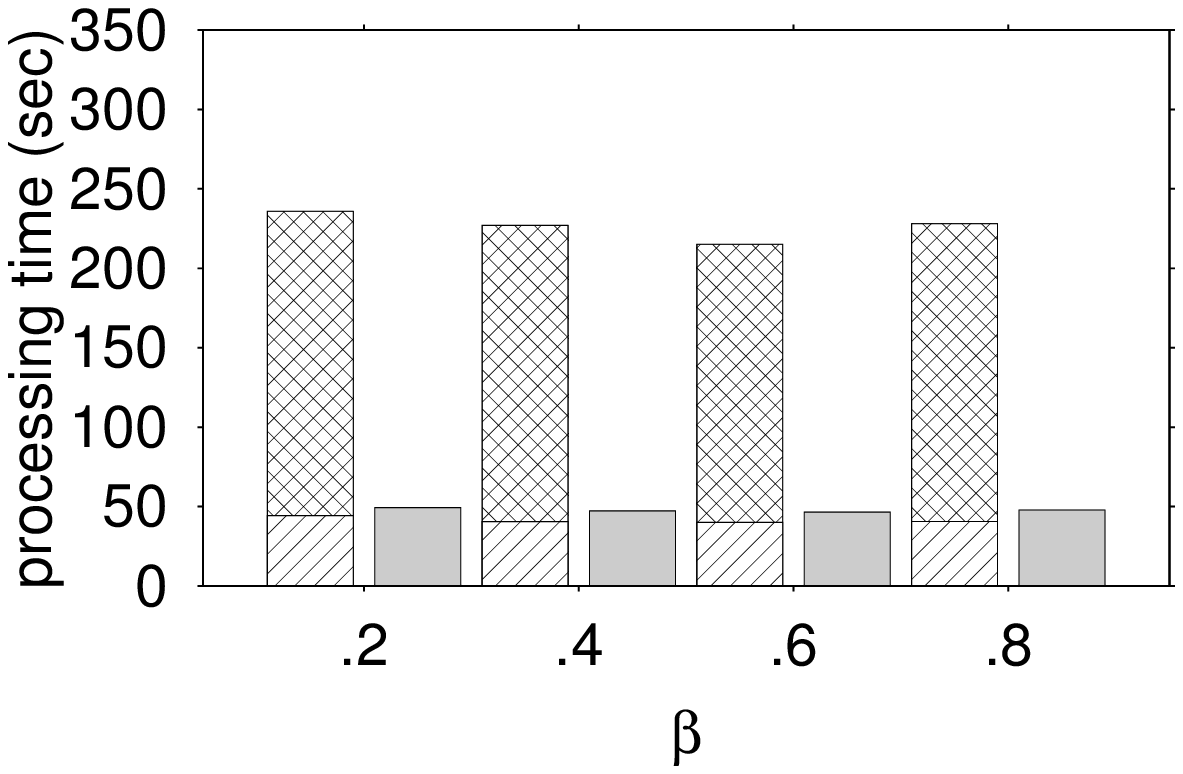}
  \vspace{-10pt}
  \caption{Processing time for the Wikipedia dataset over various configurations.}
  \label{fig:wiki:time}

  \centering
  \includegraphics[width=.18\textwidth]{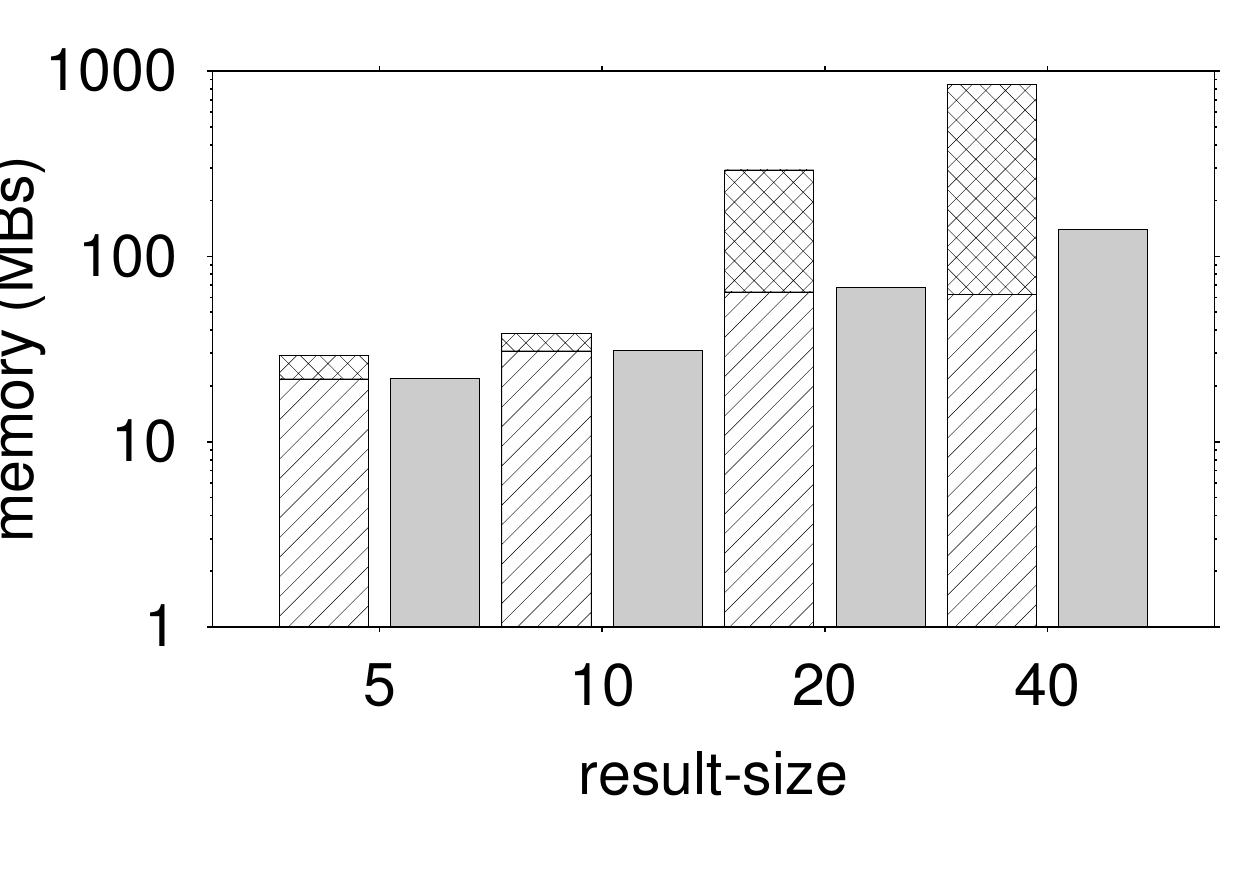}
  \includegraphics[width=.18\textwidth]{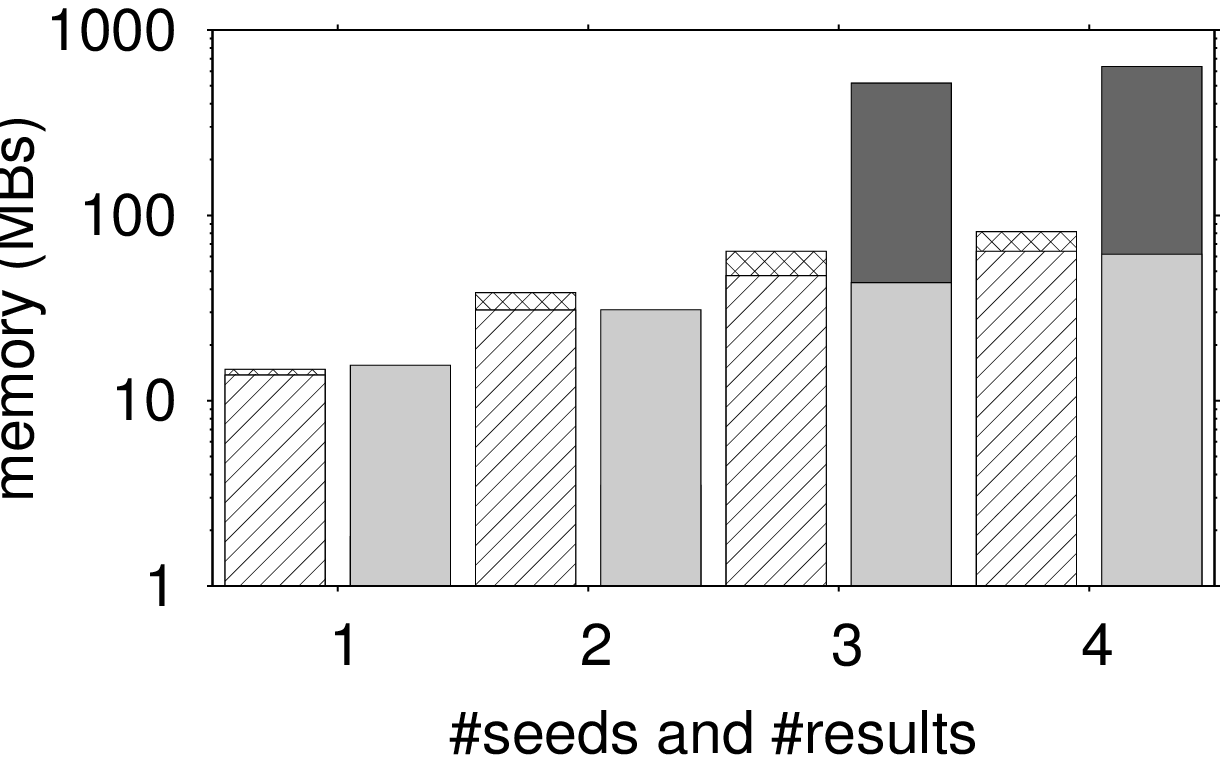}
  \includegraphics[width=.18\textwidth]{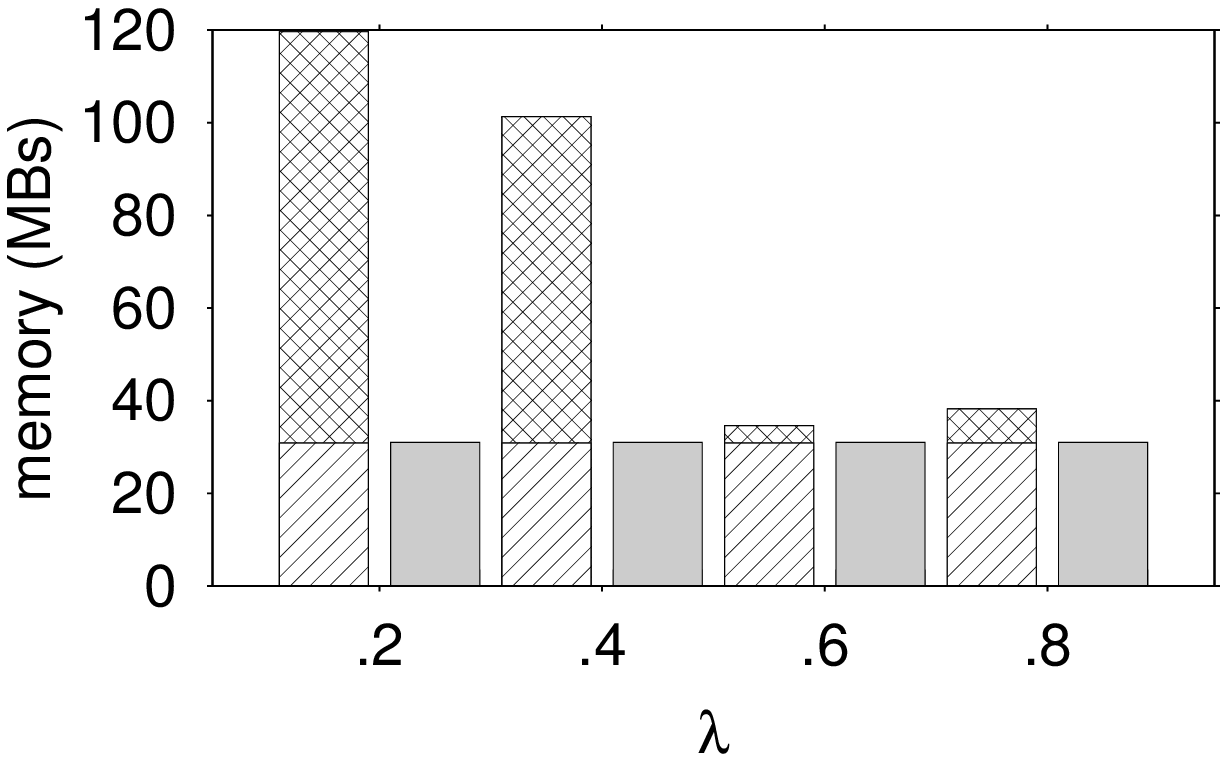}
  \includegraphics[width=.18\textwidth]{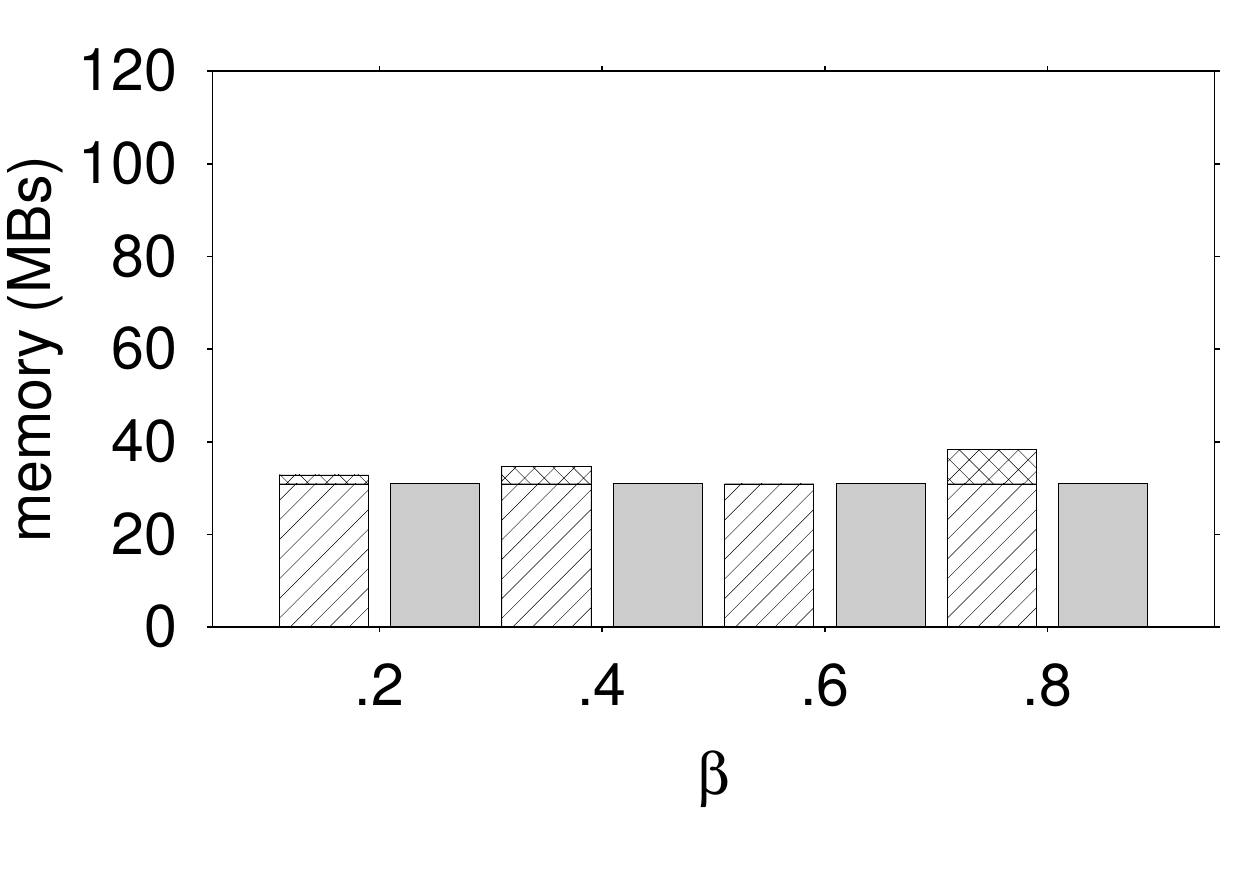}
  \vspace{-10pt}
  \caption{Allocated memory for the Wikipedia dataset over various configurations.}
  \label{fig:wiki:mem}

  \centering
  \includegraphics[width=.18\textwidth]{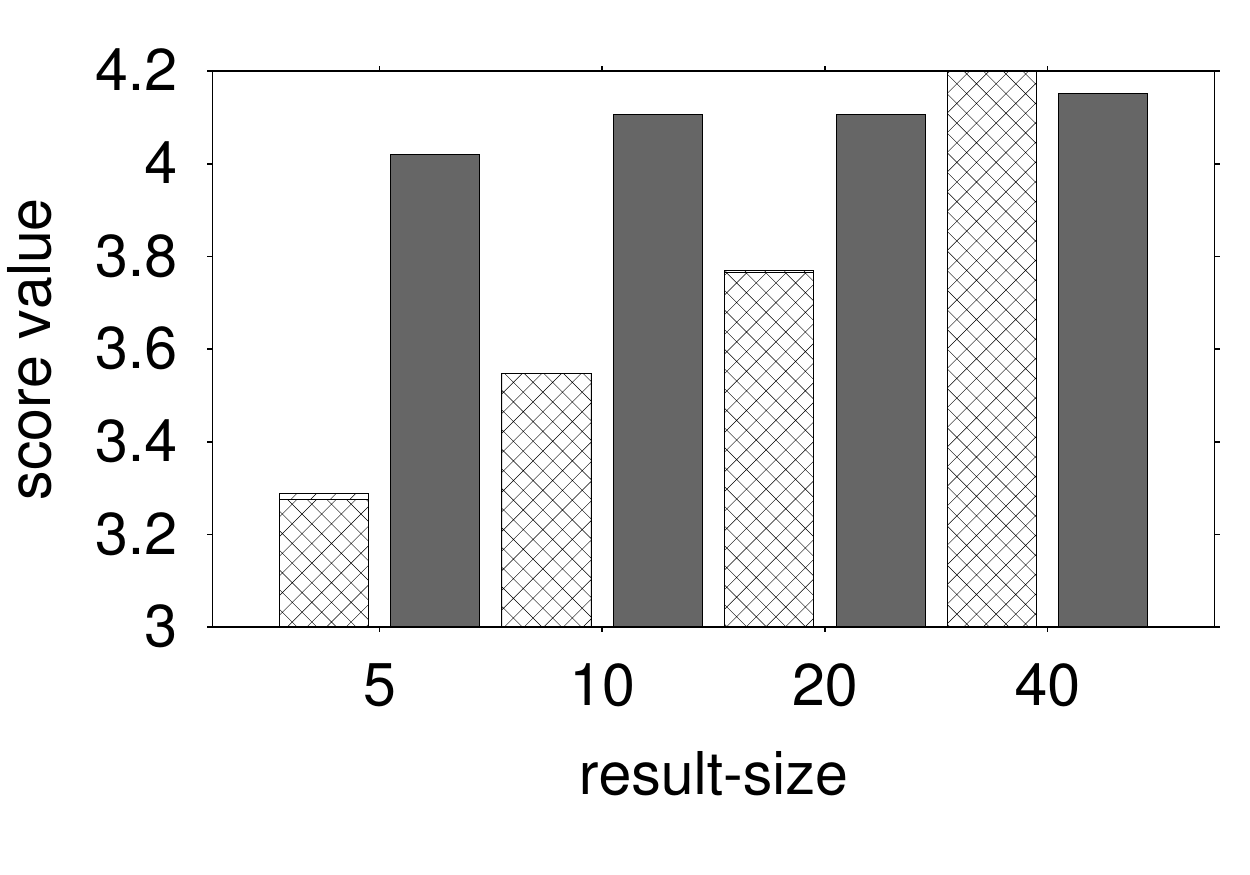}
  \includegraphics[width=.18\textwidth]{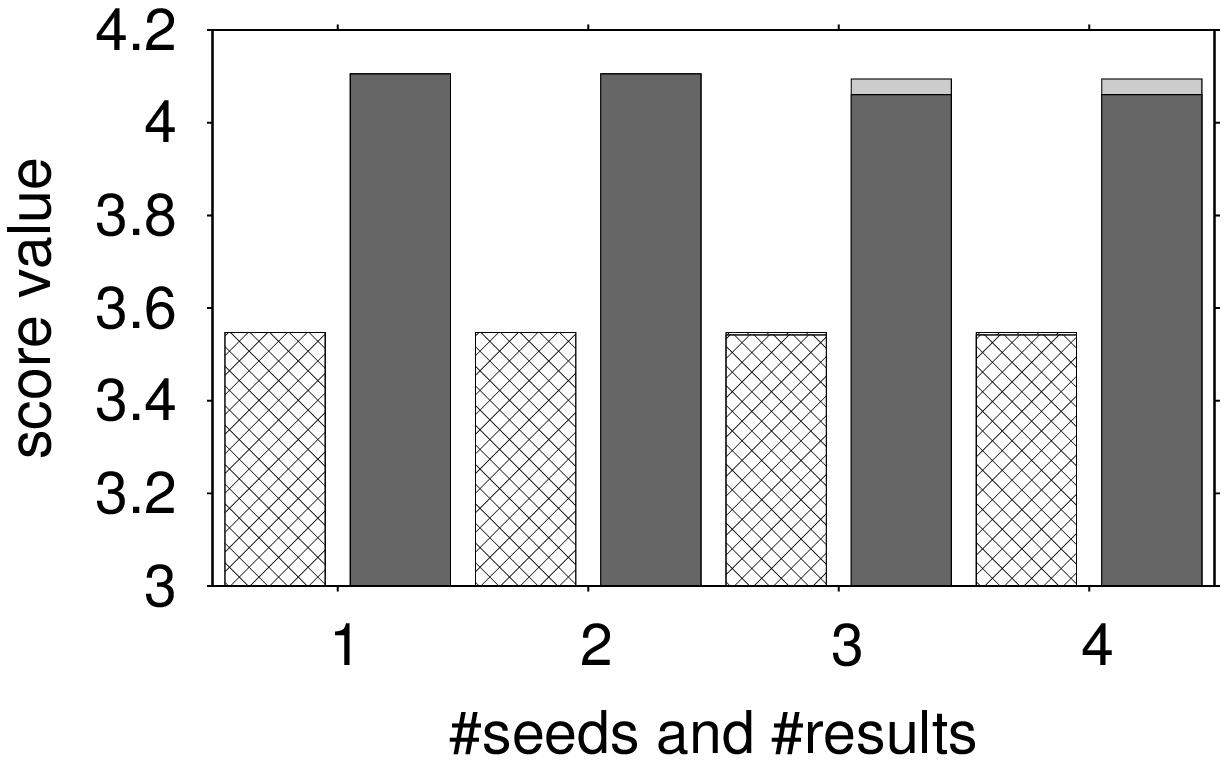}
  \includegraphics[width=.18\textwidth]{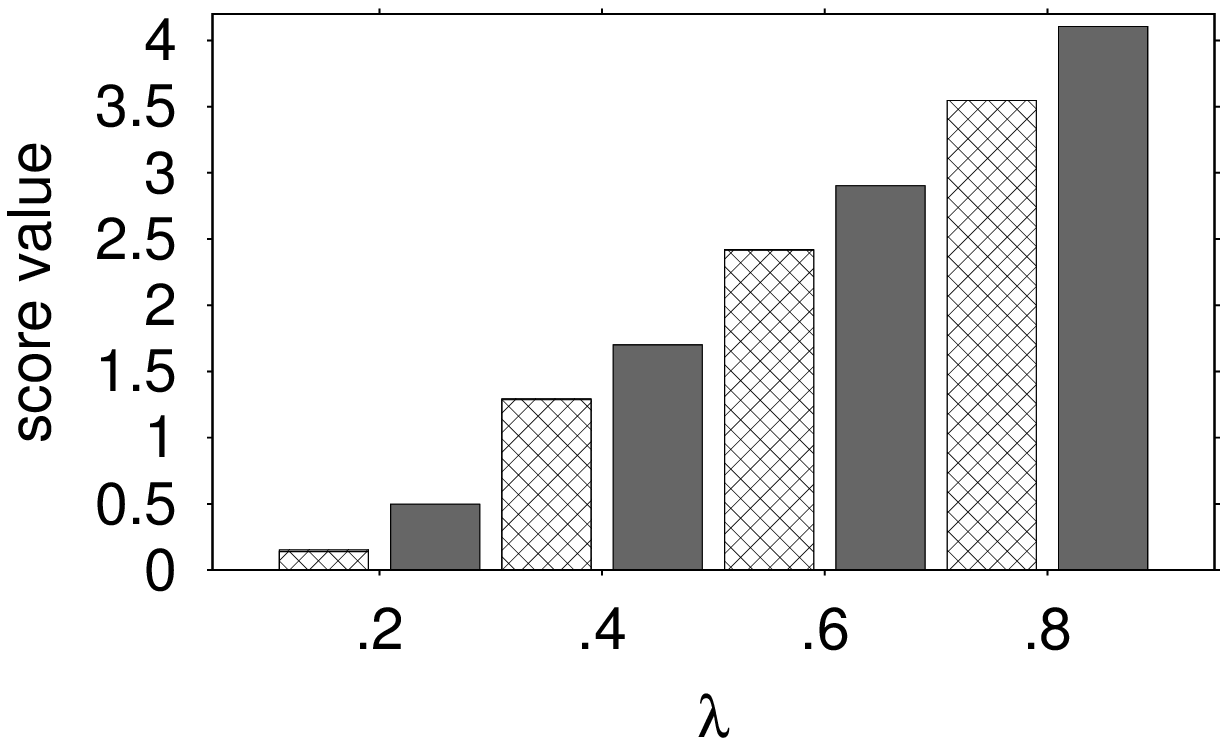}
  \includegraphics[width=.18\textwidth]{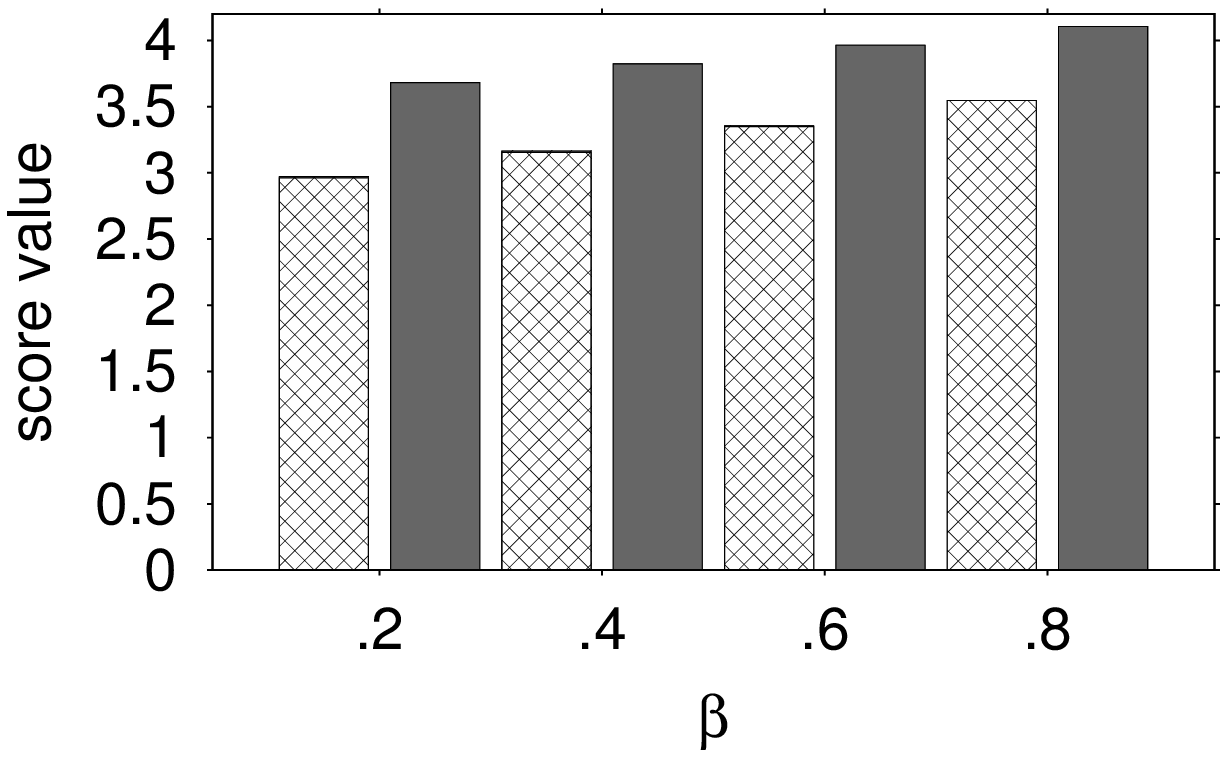}
  \vspace{-10pt}
  \caption{Rank function scores for the Wikipedia dataset over various configurations.}
  \label{fig:wiki:score}
\end{figure*}

In Figures \ref{fig:synth:time}, \ref{fig:synth:mem} and \ref{fig:synth:score}, we 
present our results for the synthetic workload, whereas in Figs.~\ref{fig:wiki:time}, 
\ref{fig:wiki:mem} and \ref{fig:synth:score} we show our results for the Wikipedia 
dataset. 
First and foremost, a prevalent observation throughout our evaluation is that our methods 
clearly outperform the \texttt{BestCoverage} method from \cite{KucuktuncSKC13} in terms of all metrics, 
processing time, allocated memory, and result quality, for all configurations of our 
parameters. In particular, our paradigm is faster by several orders of magnitude and 
consumes significantly less memory. More importantly, unlike our competitor, our approach 
is not limited to examining web-pages up to a particular distance, and it therefore 
returns a result of better quality. The main reason their performance deteriorates 
lies with the fact that they do not traverse the graph carefully as they do not exploit 
the information originated from previous traversals. For instance, they consider web-pages 
up to a certain number of hops away from the elements of the current result, but still, 
the cost of computing the score of a candidate document remains high, as our evaluation 
confirms, since many other documents can intervene in between when trying to compute the 
network distance between the candidate document and every other result item. On the other 
hand, our scheme performs a coordinated search around each result item and the query center 
until certain constraints are satisfied. What is more, according to our scheme we do not 
have to traverse any subpath twice, not even for score computation. Remember how we expand 
each distance-heap and insert the new node at the respective score-heap until it is encountered 
from all sources. That is the moment it becomes a candidate result. Never again any part of 
the path to that node from any source is visited. In particular, we reckon that graph 
data have specific characteristics that should be carefully considered in order to design 
effective and efficient processing methods. We display the results for our competitor for 
the synthetic workload only.

\textbf{Effect of collection size $|V|$.}
We can study the impact of the number of documents in the collection from the first column 
of Figs.~\ref{fig:synth:time}, \ref{fig:synth:mem} and \ref{fig:synth:score}. Regarding 
execution time, we observe in Fig.~\ref{fig:synth:time}(a) that the first phase of our 
paradigm, consisting of our heuristic method, requires less time than our hill-climbing 
approach. The first phase requires almost just as much time for both \emph{min-avg} and 
\emph{min-max} ranking criteria. However, more time is spent during the latter phase when 
\emph{min-avg} ranking criteria are used. We observe the same pattern in Fig.~\ref{fig:synth:mem}(a) 
where the amount of memory allocated by each method is shown. In Fig.~\ref{fig:synth:score}(a), 
the result quality seems to be only slightly affected by the size of the collection, and just 
for the heuristic method under \emph{min-max} ranking criteria. Notably, for \emph{min-sum} 
ranking the number of the web-documents has no effect at all.

\begin{figure}[hb]
  \centering
  \includegraphics[width=.4\textwidth]{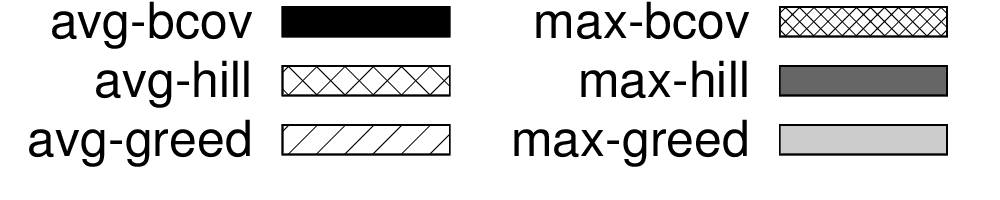}
  \vspace{-8pt}
  \caption{Legend for compared methods and ranking criteria.}
  \label{fig:legend}
\end{figure}

\textbf{Effect of number of hyper-links $|E|$.}
The second column illustrates performance in terms of the number of hyper-links to other 
documents of the collection. Apparently, processing time grows with the number of links 
per document in Fig.~\ref{fig:synth:time}(b). The reason behind this behavior lies with 
the fact that as the size of the collection remains the same and equal to the default value, 
and increasing the fan-out parameter brings the documents closer to the query center. Subsequently, 
we expect more documents to be examined now in order to ascertain whether they are qualified 
to be added to the result, and more documents will constitute candidate result items to be 
compared with other eligible ones. This imposes a computational overhead that is reflected 
in Fig.~\ref{fig:synth:time}(b) accordingly. Fig.~\ref{fig:synth:mem}(b), the memory requirements 
increase for \emph{min-max} criteria as expected, but especially for \emph{min-avg} criteria 
allocated memory seems a little irregular with the fan-out degree. What could exactly cause an 
increase in the allocated memory? First, for our hill-climbing approach that would be the number 
of qualified subsets waiting to be further processed. Keep in mind that with each such subset
we construct an iterator that allows us to traverse the best items to complement that particular 
subset. Evidently, the number of subsets that are kept simultaneously increases with the fan-out 
degree for \emph{min-max}. However, for \emph{min-avg} their number increases for the first two 
configurations to gradually diminish later on. In practice, the \emph{min-avg} ranking function 
allows for more effective pruning as it enables ranking of greater granularity. More specifically, 
as the cardinality of the result-set increases, and has already been filled with well-diversified 
objects that constitute good representatives around the query center, the remaining vertices 
achieve scores that are almost identical regardless where they lie in the graph, for they will 
always have an item of the result-set somewhere relatively close, at least no further than any 
other vertex in the graph. We reckon that this is the critical value of the cardinality of the 
result-set for \emph{min-max} ranking criteria, and this situation can be used as a termination 
criterion for the method when encountered. This is the point beyond which there is no practical 
distinction for that particular query center among the vertices that have been excluded from the 
result-set. This is also the main reason that more candidate subsets accumulate, something which 
is reflected on the allocated memory. This intricate detail explains most of the discrepancies in 
the behavior of the two ranking functions. Second, we run a configuration for our heuristic scheme 
where we limit the number of candidates by $k_g$, with a separate seed starting from each of the 
most similar documents to the query center, and therefore, the memory requirements are significantly 
less. 

\textbf{Effect of result size $n$.}
The third column illustrates performance in terms of the result size. Apparently, execution 
time grows very fast with the cardinality of the result-set in Fig.~\ref{fig:synth:time}(c) 
for both phases, even though the latter phase is more expensive under \emph{min-avg} ranking 
criteria. In Fig.~\ref{fig:synth:mem}(c), the memory allocated during the latter phase is 
increased compared to the amount of allocated memory during the first phase. Again, that gap 
is increased for \emph{min-avg} ranking criteria. In Fig.~\ref{fig:synth:score}(c), we see the 
scores deteriorate with $n$, but this is only attributed to the augmentation of the result with 
more elements. Since those are selected incrementally, less relevant elements start to appear 
whose dissimilarity from the rest of the result items evidently fails to compensate for that 
relevance loss.

\textbf{Effect of number of seeds $k_g$ and results $k_c$.}
We depict in the fourth column the performance in terms of the number of seeds produced 
during the first phase, and the number of results returned by the second phase from 
which we select the best to return to the user. This is an important system parameter 
of our scheme that does not affect our competitor, and therefore, no results are shown 
for the \texttt{BestCoverage} method here. We also remind the reader that each of those 
results has cardinality defined according to the aforementioned parameter $n$. Time grows 
linearly in Fig.~\ref{fig:synth:time}(d) for \emph{min-avg} ranking criteria and sub-linearly 
for \emph{min-max} ranking criteria. We make identical remarks in Fig.~\ref{fig:synth:mem}(d) 
where the memory allocated by our heuristic scheme increases linearly, whereas seems to grow 
faster for our hill-climbing approach. In Fig.~\ref{fig:synth:score}(d), we see the beneficial 
effect of this parameter where the quality of the score ameliorates. This is expected due to 
the wider variety of sets from which we select the final result. However, the improvement 
is not surprising and it indicates that even for small values of this parameter our 
paradigm is in position of constructing a well-diversified set. Subsequently, we claim 
that the time overhead stemming from increasing this parameter to large values does not 
compensate for the improvement in the quality of the result for our synthetic dataset. 
Setting this parameter to a small value (say two or three, just like we do for the default 
value) is sufficient.

\textbf{Effect of number of lemmas $|W|$.}
The fifth column represents our results with regard to the number of lemmas observed per 
web-document. For our configuration the lemma frequencies follow a zipfian distribution 
with skewness factor equal to $0.1$. Processing time and the amount of allocated memory 
grow linearly with the number of lemmas in Figs.~\ref{fig:synth:time}(e) and \ref{fig:synth:mem}(e) 
as expected. In particular, we see in Fig.~\ref{fig:synth:mem}(e) that our heuristic 
method requires the exact amount of memory regardless the increase in the number of lemmas. 
This can be attributed to the fact that this measurement does not account for storing the 
textual descriptions themselves, which of course would now are larger and more cumbersome 
(for computing inner products for similarities, etc.), as the response times indicate, but 
only accounts for the references to those documents that each method creates. Nevertheless, 
we discern an irregular pattern for our second phase. These fluctuations testify the sensitivity 
of the latter method to this parameter. In Fig.~\ref{fig:synth:score}(e), the scores for 
\emph{min-max} ranking criteria tend to converge with the increase in the number of lemmas, 
but there is a steady advantage for \emph{min-avg} ranking. The reasons for this behavior 
lie again with the enhanced granularity that the \emph{min-avg} ranking criteria offer.

\textbf{Effect of trade-off parameters $\lambda$ and $\beta$.}
Regarding the smoothing parameters $\lambda$ and $\beta$ of the ranking function, 
those have little impact on either execution time, or allocated memory in Figs.~\ref{fig:synth:time}(f), 
\ref{fig:synth:time}(g), \ref{fig:synth:mem}(f), \ref{fig:synth:mem}(g), for either 
\emph{min-avg}, or \emph{min-max} criteria. However, this is not true for the scores 
achieved in Figs.~\ref{fig:synth:score}(f) and \ref{fig:synth:score}(g). In essence, 
applying the hill-climbing approach right after the heuristic method, improves the 
greedy result in a way that the heuristic method could not do. To elaborate, unlike 
our hill-climbing approach, the selection of the next element for the greedy method 
takes place in accordance with the previously selected items, without being able to 
undo any of the previous selections and replacing them with a better object given the 
selection of a subsequent element. The last row of our results signifies exactly the 
contribution of the latter phase to the refinement of the result. Arguably, for some 
configurations however, when the result of the first phase is good enough, the latter 
phase is not necessary, as the sets used to seed it do not lead to a subsequent better 
result for they are already well-diversified. Another way to interpret such a behavior 
would be that for those configurations, the form of the search space impedes the 
retrieval of better replacements, and therefore, search  has to be initiated from 
more seeds so as to overcome this difficulty. Also, we might want to put into the 
same category the case where the improvement does not compensate for the time overhead. 
But when can we claim that the second phase is redundant, or has nothing contribute? 
In order to answer this question a closer look is crucial so as to detect those cases. 
Apparently, the only cases where the contribution of the latter phase is subliminal is 
when either $\lambda$ or $\beta$ take low values. For every other configuration, we see 
that at some degree our hill-climbing approach succeeds in ameliorating the quality of 
the result of the heuristic method. Keep in mind that at all times $\alpha$ is set to 
zero, and thus, relevance is computed by textual similarity only. In effect, when 
$\beta$ also takes low values, then diversified ranking relies substantially on the 
content of the web-documents and their network distances play a secondary (but still 
important) role in the computation of the result. Thereby, when  both factors of the 
ranking functions (for relevance and result items' dissimilarity) agree on what aspect 
of the data to leverage, then the heuristic method leaves no room for the second phase 
to improve the result. We observe the same phenomenon for low $\lambda$-values. This 
is when our search is oriented towards result items' dissimilarity instead of their 
relevance, and since the default value of $\beta$ aggregating result items' dissimilarity 
is directed towards elements' network distances by convention, we tacitly compare against 
the case where our main concern is biased towards network characteristics in terms of 
relevance and result items' dissimilarity also. Consequently, we can rely on the heuristic 
result exclusively when both factors agree on what aspect of the data should be put emphasis.

\textbf{Additional remarks on the Wikipedia dataset.}
We show in Figs.~\ref{fig:wiki:time}, \ref{fig:wiki:mem} and \ref{fig:wiki:score} our results 
with regard to the result-size (cardinality $n$), the number of seeds $k_g$ and results $k_c$, 
but also the trade-off parameters of the used MMR ranking functions $\lambda$ and $\beta$, 
respectively. Clearly, there is a one to one association of our observations for the synthetic 
workload to the Wikipedia dataset throughout. The results from the two different workloads do 
not contradict for any configuration. Nevertheless, in the two last columns one notices that 
the hill-climbing method, corresponding to the second processing phase, cannot improve the 
intermediate result that the heuristic method produces during the first processing step. More 
specifically, we conclude that the default number of seeds (two in our case) is not sufficient 
for the second phase to find good replacements for the heuristic set collection. Notably however, 
when it does improve the heuristic intermediate result for larger $k_g$ and $k_c$ values, the 
computation overhead becomes preventive mainly to the large size of the article collection, and 
as a result does not seem to compensate for the improvement. 

Moreover, it is hard to miss the sudden increase in memory requirements for low $\lambda$ values 
in Fig.~\ref{fig:wiki:mem}(c) during the latter processing phase (hill-climbing) for \emph{min-avg} 
ranking. This configuration denotes that we are interested in dissimilar result-items mainly, rather 
than relevant. It looks like the competition for which subset will prevail is fierce as pruning 
becomes not as helpful, and this is why we have to keep the structures associated with so many 
subsets waiting to be refined and augmented with even better replacements. Remarkably, this spiking 
behavior is not present in Fig.~\ref{fig:wiki:time}(c), and this can mean only one thing. Even though 
there is an increased number of subsets to be processed, each of them  has no more than just a few 
replacements eventually that can complement them, and thus, refinement is fast for each of them.
In effect, since the default aggregation of content dissimilarity and network distance gives priority 
to the latter, graph search becomes quite limited and indicates a few good replacements within a 
well-bounded search space. Most importantly, owing to our \texttt{verso} method, this is the lossless 
way to bound the search space, whereas our competitor uses an artificial threshold of a fixed maximum 
number of hops that graph search can expand, even when all the interesting documents that achieve high 
scores are just a little further away from the query center. This is the most prominent advantage of 
our paradigm.

\vspace{-5pt}
\section{Conclusions}
\label{sec:concl}

To recapitulate, in this paper we studied the problem of result diversification 
over interconnected document collections, e.g. web-pages. To the best of our 
knowledge our paradigm constitutes the very first approach to consider the network 
structure formed by the documents' links apart from their content.
Our solution relies on greedy heuristics to obtain an initial set of candidate 
solutions, refined through a hill-climbing approach. The criteria we adopt are 
based on Maximal Marginal Relevance (MMR) ranking functions which were expanded 
for our purposes so as to compromise objects' network distances with their 
content similarities according to their textual descriptions. Withal, special 
parameters can be used to configure the trade-off of one feature over the other, 
in such a way that the users' preferences can be accommodated to the utmost degree. 
Several optimizations have been employed in our paradigm so as to make feasible 
the processing of large graphs. Moreover, our elegant design allows for the 
tweaking of the response times when required, sometimes at a loss over the result 
quality if necessary. Finally, a thorough and meticulous experimental evaluation 
validates the effectiveness and efficiency of our method. Last but not least, we 
launched a prototype implementing our methods. It runs over the Wikipedia dataset 
but can be extended with other document collections, as well.

\vspace{-5pt}
\bibliography{diversion}

\begin{thebibliography}{10}

\bibitem{AgrawalGHI09}
R.~Agrawal, S.~Gollapudi, A.~Halverson, and S.~Ieong.
\newblock Diversifying search results.
\newblock In {\em WSDM}, pages 5--14, 2009.

\bibitem{CalinescuCPV11}
G.~C{\u{a}}linescu, C.~Chekuri, M.~P{\'{a}}l, and J.~Vondr{\'{a}}k.
\newblock Maximizing a monotone submodular function subject to a matroid
  constraint.
\newblock {\em {SIAM} J. Comput.}, 40(6):1740--1766, 2011.

\bibitem{CarbonellG98}
J.~G. Carbonell and J.~Goldstein.
\newblock The use of {MMR}, diversity-based reranking for reordering documents
  and producing summaries.
\newblock In {\em SIGIR}, pages 335--336, 1998.

\bibitem{ClarkeKCVABM08}
C.~L.~A. Clarke, M.~Kolla, G.~V. Cormack, O.~Vechtomova, A.~Ashkan,
  S.~B{\"u}ttcher, and I.~MacKinnon.
\newblock Novelty and diversity in information retrieval evaluation.
\newblock In {\em SIGIR}, pages 659--666, 2008.

\bibitem{DrosouP10}
M.~Drosou and E.~Pitoura.
\newblock Search result diversification.
\newblock {\em SIGMOD Record}, 39(1):41--47, 2010.

\bibitem{disc1}
M.~Drosou and E.~Pitoura.
\newblock Disc diversity: result diversification based on dissimilarity and
  coverage.
\newblock {\em {PVLDB}}, 6(1):13--24, 2012.

\bibitem{disc2}
M.~Drosou and E.~Pitoura.
\newblock Multiple radii disc diversity: Result diversification based on
  dissimilarity and coverage.
\newblock {\em {ACM} Trans. Database Syst.}, 40(1):4, 2015.

\bibitem{topKsigmod}
P.~Fraternali, D.~Martinenghi, and M.~Tagliasacchi.
\newblock Top-k bounded diversification.
\newblock In {\em SIGMOD}, pages 421--432, 2012.

\bibitem{GollapudiS09}
S.~Gollapudi and A.~Sharma.
\newblock An axiomatic framework for result diversification.
\newblock {\em IEEE Da. Eng. Bul.}, 32(4):7--14, 2009.

\bibitem{Haritsa09}
J.~R. Haritsa.
\newblock The {KNDN} problem: A quest for unity in diversity.
\newblock {\em IEEE Data Eng. Bull.}, 32(4):15--22, 2009.

\bibitem{HeTMS12}
J.~He, H.~Tong, Q.~Mei, and B.~K. Szymanski.
\newblock Gender: {A} generic diversified ranking algorithm.
\newblock In {\em NIPS 2012}, pages 1151--1159, 2012.

\bibitem{KucuktuncSKC13}
O.~K{\"{u}}{\c{c}}{\"{u}}ktun{\c{c}}, E.~Saule, K.~Kaya, and {\"{U}}.~V.
  {\c{C}}ataly{\"{u}}rek.
\newblock Diversified recommendation on graphs: pitfalls, measures, and
  algorithms.
\newblock In {\em {WWW} 2013}, pages 715--726, 2013.

\bibitem{LiY13}
R.~Li and J.~X. Yu.
\newblock Scalable diversified ranking on large graphs.
\newblock {\em {IEEE} Trans. Knowl. Data Eng.}, 25(9):2133--2146, 2013.

\bibitem{LiuSC09}
Z.~Liu, P.~Sun, and Y.~Chen.
\newblock Structured search result differentiation.
\newblock {\em PVLDB}, 2(1):313--324, 2009.

\bibitem{MeiGR10}
Q.~Mei, J.~Guo, and D.~R. Radev.
\newblock Divrank: {T}he interplay of prestige and diversity in information
  networks.
\newblock In {\em {SIGKDD} 2010}, pages 1009--1018, 2010.

\bibitem{NemhauserWF78}
G.~L. Nemhauser, L.~A. Wolsey, and M.~L. Fisher.
\newblock An analysis of approximations for maximizing submodular set functions
  - {I}.
\newblock {\em Math. Program.}, 14(1):265--294, 1978.

\bibitem{ilprints422}
L.~Page, S.~Brin, R.~Motwani, and T.~Winograd.
\newblock The pagerank citation ranking: Bringing order to the web.
\newblock Technical Report 1999-66, Stanford InfoLab, November 1999.

\bibitem{TongHWKL11}
H.~Tong, J.~He, Z.~Wen, R.~Konuru, and C.~Lin.
\newblock Diversified ranking on large graphs: {A}n optimization viewpoint.
\newblock In {\em {SIGKDD} 2011}, pages 1028--1036, 2011.

\bibitem{phdworkshop}
G.~Tsatsanifos.
\newblock The tantalizing new prospect of index-based diversified retrieval.
\newblock In {\em {SIGMOD/PODS 2013} Ph.D. Symposium}, pages 49--54, 2013.

\bibitem{ripple}
G.~Tsatsanifos, D.~Sacharidis, and T.~Sellis.
\newblock {RIPPLE:} {A} scalable framework for distributed processing of rank
  queries.
\newblock In {\em {EDBT} 2014}, pages 259--270, 2014.

\bibitem{VeeSA09}
E.~Vee, J.~Shanmugasundaram, and S.~Amer-Yahia.
\newblock Efficient computation of diverse query results.
\newblock {\em IEEE Data Eng. Bull.}, 32(4):57--64, 2009.

\bibitem{marios}
M.~Vieira, H.~Razente, M.~Barioni, M.~Hadjieleftheriou, D.~Srivastava, C.~Jr.,
  and V.~Tsotras.
\newblock On query result diversification.
\newblock In {\em ICDE}, pages 1163--1174, 2011.

\bibitem{XuY08}
Y.~Xu and H.~Yin.
\newblock Novelty and topicality in interactive information retrieval.
\newblock {\em JASIST}, 59(2):201--215, 2008.

\bibitem{YuLA09}
C.~Yu, L.~V.~S. Lakshmanan, and S.~Amer-Yahia.
\newblock It takes variety to make a world: {D}iversification in recommender
  systems.
\newblock In {\em EDBT}, pages 368--378, 2009.

\bibitem{ZhaiCL03}
C.~Zhai, W.~W. Cohen, and J.~D. Lafferty.
\newblock Beyond independent relevance: {M}ethods and evaluation metrics for
  subtopic retrieval.
\newblock In {\em SIGIR}, pages 10--17, 2003.

\bibitem{ZhangH08}
M.~Zhang and N.~Hurley.
\newblock Avoiding monotony: {I}mproving the diversity of recommendation lists.
\newblock In {\em RecSys}.

\bibitem{ZhangCM02}
Y.~Zhang, J.~P. Callan, and T.~P. Minka.
\newblock Novelty and redundancy detection in adaptive filtering.
\newblock In {\em SIGIR}, pages 81--88, 2002.

\bibitem{ZhuGGA07}
X.~Zhu, A.~B. Goldberg, J.~V. Gael, and D.~Andrzejewski.
\newblock Improving diversity in ranking using absorbing random walks.
\newblock In {\em HLT 2007}, pages 97--104, 2007.

\bibitem{ZieglerMKL05}
C.-N. Ziegler, S.~M. McNee, J.~A. Konstan, and G.~Lausen.
\newblock Improving recommendation lists through topic diversification.
\newblock In {\em WWW}, pages 22--32, 2005.

\end{thebibliography}
\bibliographystyle{abbrv}

\end{document}